\documentclass[final,authoryear,3p,times,twocolumn]{elsarticle}

\usepackage{epsfig}

\usepackage{amssymb}

\usepackage{lineno}
\usepackage{rotating}
\usepackage[bookmarks = true, bookmarksnumbered = true, pdfpagemode =None, pdfstartview = FitH, pdfpagelayout = SinglePage, colorlinks = true, urlcolor = blue, citecolor = monbleu]{hyperref}

\newcommand{\cm}{cm$^{-1}$ }

\journal{Icarus}

\begin{document}

\begin{frontmatter}



\title{The Origin of Nitrogen on Jupiter and Saturn from the $^{15}$N/$^{14}$N Ratio}


\author[ox]{Leigh N. Fletcher}
\ead{fletcher@atm.ox.ac.uk}
\author[swri]{T.K. Greathouse}
\author[jpl]{G.S. Orton}
\author[ox]{P.G.J. Irwin}
\author[om]{O. Mousis}
\author[ox]{J.A. Sinclair}
\author[ox]{R.S. Giles}

\address[ox]{Atmospheric, Oceanic \& Planetary Physics, Department of Physics, University of Oxford, Clarendon Laboratory, Parks Road, Oxford, OX1 3PU, UK}
\address[swri]{Southwest Research Institute, Division 15, 6220 Culebra Road, San Antonio, Texas 78228, USA}
\address[jpl]{Jet Propulsion Laboratory, California Institute of Technology, 4800 Oak Grove Drive, Pasadena, CA, 91109, USA}
\address[om]{Universit\'{e} de Franche-Comt\'{e}, Institut UTINAM, CNRS/INSU, UMR 6213, Besan\c{c}on, Cedex, France}

\begin{abstract}

The Texas Echelon cross Echelle Spectrograph (TEXES), mounted on NASA's Infrared Telescope Facility (IRTF), was used to map mid-infrared ammonia absorption features on both Jupiter and Saturn in February 2013.  Ammonia is the principle reservoir of nitrogen on the giant planets, and the ratio of isotopologues ($^{15}$N/$^{14}$N) can reveal insights into the molecular carrier (e.g., as N$_2$ or NH$_3$) of nitrogen to the forming protoplanets, and hence the source reservoirs from which these worlds accreted.  We targeted two spectral intervals (900 and 960 \cm) that were relatively clear of terrestrial atmospheric contamination and contained close features of $^{14}$NH$_3$ and $^{15}$NH$_3$, allowing us to derive the ratio from a single spectrum without ambiguity due to radiometric calibration (the primary source of uncertainty in this study).  We present the first ground-based determination of Jupiter's $^{15}$N/$^{14}$N ratio (in the range from $1.4\times10^{-3}$ to $2.5\times10^{-3}$), which is consistent with both previous space-based studies and with the primordial value of the protosolar nebula.  On Saturn, we present the first upper limit on the $^{15}$N/$^{14}$N ratio of no larger than $2.0\times10^{-3}$ for the 900-\cm channel and a less stringent requirement that the ratio be no larger than $2.8\times10^{-3}$ for the 960-\cm channel ($1\sigma$ confidence).  Specifically, the data rule out strong $^{15}$N-enrichments such as those observed in Titan's atmosphere and in cometary nitrogen compounds.  To the extent possible with ground-based radiometric uncertainties, the saturnian and jovian $^{15}$N/$^{14}$N ratios appear indistinguishable, implying that $^{15}$N-enriched ammonia ices could not have been a substantial contributor to the bulk nitrogen inventory of either planet.  This result favours accretion of primordial N$_2$ on both planets, either in the gas phase from the solar nebula, or as ices formed at very low temperatures.  Finally, spatially-resolved TEXES observations are used to derive zonal contrasts in tropospheric temperatures, phosphine and $^{14}$NH$_3$ on both planets, allowing us to relate thermal conditions and chemical compositions to phenomena observed at visible wavelengths in 2013 (e.g., Jupiter's faint equatorial red coloration event and wave activity in the equatorial belts, plus the remnant warm band on Saturn following the 2010-11 springtime storm).  

\end{abstract}

\begin{keyword}
Jupiter \sep Saturn \sep Atmospheres, composition

\end{keyword}

\end{frontmatter}


\section{Introduction}
\label{intro}

The bulk elemental and isotopic composition of a giant planet reflects the balance of chemical species in their primordial source reservoirs.  The core-nucleated accretion model of planet formation  \citep[e.g.,][]{80mizuno, 93lissauer, 89pollack, 96pollack} suggests that the giant planets grew from solid ice/rock cores by accretion of gases and solids, and then by runaway accretion of the remaining nebula gases until the feeding zone resources were exhausted \citep[e.g., by either nebula dissipation or tidal gap opening, see the recent review by][]{13helled}. The availability of planetary building blocks (metals, oxides, silicates, ices) varied with position and time within the original nebula, from refractories in the warm inner nebula to a variety of ices of water, CH$_4$, CO, CO$_2$, NH$_3$, N$_2$ and other simple molecules in the cold outer nebula.  Furthermore, the planets were not fixed within the disc, and both inward and outward migration during their evolution could have provided access to different material reservoirs at different epochs.    In this study, we use measurements of the ratio of two nitrogen isotopes ($^{15}$N/$^{14}$N) in giant planet ammonia to explore the origins of Jupiter and Saturn.  The elemental and isotopic composition `frozen into' the planetary building blocks and nebula gases was embedded into the giant planet atmosphere and preserved against subsequent re-equilibriation during the planet's evolution, providing a window onto conditions in the early solar system at the time and location of planetary formation.

Molecular nitrogen (N$_2$) is the primary reservoir of nitrogen in interstellar clouds \citep{93vandishoeck}, so one might expect it to be the dominant form for nitrogen delivery to forming solar system objects \citep{80lewis}.  However, although N$_2$ is present in the atmospheres of Venus (3.5\% by volume), Earth (78\%), Mars (2.7\%) and Titan (98\%), this is a secondary product of dissociation of the original carrier molecule that was outgassed from the primordial interiors.  Conversely, nitrogen on the giant planets is expected to reflect the composition of the primordial solar nebula.  Giant-planet nitrogen is in the form of NH$_3$, thermochemically converted from N$_2$ in the hot deep interiors \citep{94fegley}.  Unfortunately, NH$_3$ (along with many of the most abundant chemical species, like oxygen and sulphur) is removed from the gas phase by condensation to form a series of cloud layers, locking away the volatiles (water, ammonia and H$_2$S) in the deep troposphere that is largely hidden from the capabilities of remote sensing. Nevertheless, nitrogen has been found to be enriched over solar abundances on both Jupiter and Saturn \citep{85depater, 98niemann, 99owen, 04wong, 09fletcher_ch4}, supporting the hypothesis of enrichment by accreted planetesimals during the formation process.  Accretion of nitrogen in the gas phase alone cannot produce these supersolar enrichments without some contribution of nitrogen species trapped in solids.  Further constraints are provided by the Galileo Probe Mass Spectrometer \citep[GPMS,][]{98niemann, 99atreya, 00mahaffy, 04wong}, which suggests that Jupiter's elements are uniformly enhanced by a factor of $4\pm2$ compared to protosolar abundances.  This `solar balance' of enrichment requires that all elements, particularly nitrogen and argon, were trapped just as efficiently as carbon in Jupiter's source reservoir.  

Jupiter's uniform supersolar enrichment can be explained by two different scenarios:  the accretion of volatiles trapped in the water matrix of amorphous ice planetesimals, either at large heliocentric distances or in a very cold and static disc \citep[ice formation temperatures $T_f<20$ K, e.g.,][]{99owen, 01owen, 03owen, 06owen}, or by the crystallisation of ices (clathrates and/or pure condensates) in Jupiter's feeding zone during the cooling of the protosolar nebula \citep{01gautier, 85lunine, 04hersant, 09mousis, 09mousis_titan}.  In the latter scenario, bulk enrichments are governed by the trapping efficiencies of volatiles and the formation of crystalline ices as the nebula cooled \citep[e.g.,][]{09mousis} and the condensation of pure ices; in the former scenario the balance of elements in primitive amorphous ices is governed by the cold chemistry of the interstellar medium, although the mechanism for transporting these primitive objects to the forming giant planets is unclear.  In both theories, the disc temperature in the source reservoir (which could have cooled over hundreds of kelvin at the location of giant planet formation) is crucial in establishing the dominant molecular carrier trapped in the icy building blocks - cold environmental temperatures would permit trapping of N$_2$ in ices, whereas warmer conditions would lead to deficiencies in neon, argon and N$_2$ in those icy planetesimals and favour the trapping of nitrogen compounds like NH$_3$ \citep[e.g.,][]{99owen}.   This work will discuss giant planet $^{15}$N/$^{14}$N ratios in the context of these competing theories, allowing us to distinguish the primary molecular carrier delivering nitrogen to the forming proto-Jupiter and proto-Saturn.

%

\subsection{The $^{15}$N/$^{14}$N ratio}

To ascertain the specific molecular carrier of nitrogen, we can exploit the fact that chemical fractionation in different nitrogen-bearing molecules (either during the protosolar or earlier interstellar periods) would have caused enrichment of the heavier $^{15}$N isotope in nitrogen compounds (e.g., HCN, NH$_3$, CN) compared to molecular nitrogen (N$_2$) \citep{95owen, 00terzieva}.  The degree of $^{15}$N enrichment in these compounds varies with temperature and formation time, and the value changes from compound to compound due to escape and fractionation processes.  Furthermore, fractionation by diffusive separation, ion pick-up and atmospheric sputtering caused by solar winds can aid the escape of the lighter isotope from terrestrial-type planets to create a $^{15}$N enrichment, as observed on Mars and Titan \citep{10atreya}.  Nevertheless, the lowest `primordial' $^{15}$N enrichments are expected in N$_2$, which would condense or be efficiently adsorbed onto ices only at extremely low temperatures.    The highest $^{15}$N-enrichments reported are those in cometary NH$_3$, such as the $^{15}$N/$^{14}$N ratio of $7.9^{+4.6}_{-2.6}\times10^{-3}$ from an average of spectra from 12 comets \citep{14rousselot} and the recent observation of NH$_2$ emission from Comet C/2012 S1 (ISON) by \citet{14shinnaka}, who estimate a $^{15}$N/$^{14}$N ratio $7.2^{+2.7}_{-1.6}\times10^{-3}$.  On Earth, the $^{15}$N/$^{14}$N ratio of $3.66\times10^{-3}$ is similar to that found in primitive meteorites \citep{12marty}, but much lower than that found in comets, so cometary ammonia could not have been the main supplier of Earth's nitrogen.  But could cometesimal delivery have played an important role in delivering heavy elements in the outer solar system?   Comets are notoriously deficient in N$_2$, with the bulk of cometary nitrogen being bound as compounds such as NH$_3$, HCN and CN \citep{04bockelee-morvan}.  Identifying the molecular carrier of nitrogen to Jupiter and Saturn would help to address this question.


\textit{In situ} measurements of the ratio in Jupiter by the Galileo Probe Mass Spectrometer (GPMS) suggested a $^{15}$N/$^{14}$N ratio of $(2.3\pm0.3)\times10^{-3}$, considerably lower than the terrestrial value \citep{98niemann, 00mahaffy, 01owen}.  This result was independently confirmed by mid-infrared disc-averaged spectroscopy from the Infrared Space Observatory \citep[ISO, $1.9^{+0.9}_{-1.0}\times10^{-3}$,][]{00fouchet} and spatially-resolved spectroscopy from the Cassini Composite Infrared Spectrometer \citep[CIRS, $(2.2\pm0.5)\times10^{-3}$,][]{04fouchet,04abbas}.  \citet{03atreya} point out the remarkable similarity of these values, despite the fact that the mid-IR results are pertinent to the sub-saturated regions above the cloud-tops (around 0.5 bar), whereas the probe measurements were from higher pressures (0.8-2.8 bar) in a region of anomalous 'hot-spot' meteorology.  CIRS observations \citep{04fouchet,04abbas} found no evidence for spatial variability of the ratio, indicating insensitivity to condensation fractionation.  Indeed, \citet{00fouchet, 04fouchet} carefully assessed the isotopic fractionation effects related to cloud condensation (i.e., condensation preferentially segregates the heavier isotopologue into the condensed phase, thereby depleting the vapour phase) and found it to have a negligible impact.  Furthermore, \citet{07liang} demonstrated that ammonia photolysis (which is more efficient for the heavier isotopologue) is likely to be diluted by tropospheric mixing and condensation effects.  Jupiter's low $^{15}$N enrichment favours delivery as primordial N$_2$ (either as a gas, condensate or trapped in ices), rather than delivery as primordial NH$_3$.  Furthermore, the isotopic ratio is thought to represent the primordial solar nebula value, being similar to: $^{15}$N/$^{14}$N$=(2.36\pm0.02)\times10^{-3}$ found in a meteorite inclusion \citep{07meibom}; the bulk Sun value of $(2.27\pm0.03)\times10^{-3}$ derived from solar wind samples obtained by Genesis \citep{11marty}; and the solar wind estimate of  $<2.8\times10^{-3}$ using nitrogen trapped in the outer rims of individual lunar soil grains \citep{00hashizume}.

In contrast, Saturn's $^{15}$N enrichment has never been measured.  In the absence of an \textit{in situ} probe \citep{09marty, 14mousis}, we must rely on remote sensing measurements to study Saturn's ammonia distribution.  Saturn's cold atmosphere pushes the main ammonia ice condensation cloud deeper than in Jupiter, and it is obscured by a ubiquitous haze in the upper troposphere \citep{09west}.  Ammonia follows a sub-saturated distribution above the clouds and may be converted photochemically to hydrazine (N$_2$H$_4$) in the upper troposphere.  The spatial distribution has proven difficult to obtain \citep{12hurley}, but appears to be enhanced at low latitudes by vigorous upwelling \citep{11fletcher_vims}.  The cold atmospheric temperatures and deep ammonia cloud mean that NH$_3$ has a limited influence over Saturn's mid-infrared (8-12 $\mu$m) spectrum, which is instead dominated by phosphine absorption features \citep{07fletcher_ph3}.  Indeed, the noise performance of Cassini/CIRS has proven insufficient to constrain Saturn's $^{15}$N/$^{14}$N ratio.  

A measurement of Saturn's $^{15}$N enrichment would be particularly intriguing, as the Huygens probe Gas Chromatograph and Mass Spectrometer \citep[GCMS,][]{05niemann} discovered substantial $^{15}$N enrichment in Titan's atmosphere - a $^{15}$N/$^{14}$N ratio of $(5.96\pm0.02)\times10^{-3}$ in N$_2$ \citep{10niemann}, greater than that on Earth and considerably larger than the ratio found on Jupiter.  Accretion of a $^{15}$N-enriched nitrogen compound from Saturn's feeding zone \citep[e.g., primordial NH$_3$,][]{10atreya} and subsequent production of a secondary atmosphere from the original carrier could explain Titan's $^{15}$N enrichment if fractionation during atmospheric escape cannot account for this high value (see Section \ref{discuss}).  We therefore have two different scenarios for the $^{15}$N/$^{14}$N ratio of Saturn: either $^{15}$N-enriched due to accretion from a primordial NH$_3$ reservoir like Titan (although Titan would have accreted long after the majority of Saturn's mass had been delivered), or $^{15}$N-poor due to accretion from a primordial N$_2$ reservoir like Jupiter.  Any active chemistry occurring within either the feeding zones (for the gas giant) or later sub-nebulae (for the regular satellites) would have preserved the isotopic ratio of the original source reservoirs, allowing us to measure the $^{15}$N/$^{14}$N ratio to identify these cosmochemical reservoirs.


This ground-based spectroscopic study focusses on the $^{15}$N/$^{14}$N ratio in ammonia to provide confirmation of the jovian value and the first constraints on this ratio in Saturn.  In Section \ref{data} we describe thermal infrared spectroscopy of Jupiter and Saturn obtained in February 2013 by the TEXES instrument \citep[Texas Echelon cross Echelle Spectrograph,][]{02lacy} on NASA's Infrared Telescope Facility (IRTF).  Section \ref{model} describes the radiative transfer and spectral inversion approach used to determine the tropospheric temperature, aerosol opacity and gaseous composition (PH$_3$, $^{14}$NH$_3$ and $^{15}$NH$_3$) on both planets in Section \ref{results}.  Implications of the nitrogen isotopic composition for the formation of the giant planets will be discussed in Section \ref{discuss}.

\section{TEXES Observations of Jupiter and Saturn}
\label{data}

\subsection{IRTF/TEXES Observations in 2013}

We observed the gas giants Jupiter and Saturn over a series of consecutive nights in February 2013. Details of the observing run are presented in Table \ref{tab:data}. Jupiter was 4.7 AU from Earth on February 10, 2013 and at the end of its 2012-13 apparition (opposition on December 12, 2012), appearing 41.6" in diameter at a heliocentric longitude $L_s=120^\circ$ (local northern summer). Saturn was at the start of its apparition (opposition on April 28, 2013); 9.58 AU from Earth, appearing 17.2" in diameter at $L_s=42^\circ$ (local northern spring).  The combination of the 3-m diameter primary mirror of NASA's Infrared Telescope Facility (IRTF) and the TEXES instrument \citep[Texas Echelon cross Echelle Spectrograph,][]{02lacy} provided spatially-resolved Q- (17-25 $\mu$m) and N-band (7-13 $\mu$m) spectroscopy of both targets.  Jupiter was observed to demonstrate our ability to measure the jovian $^{15}$N/$^{14}$N ratio from Earth, providing confidence in our first constraints on the saturnian value.


This study targeted pressure-broadened lines of ammonia and phosphine, so the highest spectral resolutions permitted by TEXES ($R=\lambda/\Delta\lambda\approx80000$) were not required.  By shifting to lower resolving powers ($R\approx2000-3000$) we were able to capture more of the spectrum in a single observation (a band approximately 20 cm$^{-1}$ wide), permitting simultaneous retrievals of temperature and composition from individual bands.  Two spectral settings were selected based on the following criteria:  (i) they must be suitable for $^{15}$NH$_3$ measurements on both ammonia-dominated Jupiter and phosphine-dominated Saturn; (ii) they must avoid regions of low terrestrial transmission due to H$_2$O and O$_3$; (iii) they must have both $^{15}$NH$_3$ and $^{14}$NH$_3$ features within 10-15 cm$^{-1}$ of one another to be captured simultaneously; (iv) they must avoid regions of strong hydrocarbon emission at $\lambda> 11.5 \mu$m; and (v) they should allow for an independent reconstruction of PH$_3$.  Synthetic TEXES spectra with varying ammonia and phosphine abundances between 850 and 1200 cm$^{-1}$ were generated prior to the observing run, and settings at $900\pm10$ cm$^{-1}$ and $960\pm10$ cm$^{-1}$ were selected.  The diffraction-limited spatial resolution is approximately 1 arcsec at these spectral settings.  The spectral resolving power can be estimated from the grating equation, which depends on the grating angle and the angular size of the slit and provides values of $R=2896$ (or $\Delta\nu=0.31$ cm$^{-1}$) for the 900 cm$^{-1}$ setting; and $R=2664$ (or $\Delta\nu=0.36$ cm$^{-1}$) for the 960 cm$^{-1}$ setting.  Additional settings near 1002 and 1021 cm$^{-1}$ provide a better $^{15}$NH$_3$ signature, but are completely obscured by telluric ozone absorption.  Furthermore, our observations use the same 903.1 cm$^{-1}$ P(3) $\nu_{2a}$ transition of $^{15}$NH$_3$ as observed by Cassini/CIRS for Jupiter \citep{04fouchet} as a check of our technique.   


For Jupiter observations, we employed a standard TEXES technique of aligning the slit to celestial north and scanning from east to west across the target to assemble an image over several minutes (Fig. \ref{images}).  Jupiter's rotation axis was close to the north-south direction on the sky. The 1.4" slit was moved in 0.7" increments (2-second integrations at each step) to assemble one scan, and then repeated several times to build up signal-to-noise.  The scans started and finished with blank sky spectra to permit subtraction of sky background emission from the on-source measurements.  Unlike mid-infrared imaging, we do not use the chopping secondary, and the scan mode uses off-target scan steps instead of nodded pairs.  Given the lower signal from Saturn, we primarily used TEXES in a nodding mode rather than scanning mode, nodding Saturn along the slit to produce positive and negative spectra that were later aligned and added.  Scan-mode Saturn observations were attempted in both 2012 and 2013 at medium spectral resolutions ($R\approx10000$), and on February 9th in low resolution mode (Table \ref{tab:data}) to produce the images in Fig. \ref{images}.  The medium resolution scanning observations will be used in Section \ref{sat_15N} to verify conclusions from the low-resolution nod settings.

A room-temperature black body (a high-emissivity metal chopper blade painted black just above the Dewer entrance window) was observed prior to each scan or nod to serve as both a flat field and a radiance calibrator \citep{02lacy, 11greathouse}.  To first order, we assume that the temperature of this metal plate approximates the sky and telescope temperatures, so that the differences between the black body and sky observations can be used to account for instrument and telluric emission.  Using two observations of the black body and two of the sky, we are able to both flat-field and radiometrically calibrate the target data.  The black body fills the field of view (FOV) in the same way as the extended target, eliminating the need for FOV-filling corrections when point sources are used as divisors.   Absolute calibration via this technique was found to be more efficient and reliable than observations of standard divisors \citep[mid-IR bright asteroids, for example,][]{02lacy}, permitting maximal time on the science target.  Cohen spectrophotometric standards \citep[mainly early K dwarfs with SiO spectral features in the N-band,][]{99cohen} are not suitable at high spectral resolution as their spectra contain multiple photospheric lines and are often not bright enough to provide adequate signal in a short observation.  Furthermore, the moon could not be used as a calibrator as its high temperature causes saturation, and the rapid variation of temperature over small spatial scales makes the moon a poor flat field.  Removal of telluric `sky' emission at every scan step is done by subtracting the small portion of sky observed at either end of the slit.  However, this sky subtraction does not remove telluric absorption completely (it does a good job for water and CO$_2$, but not O$_3$ as it is high and cold in Earth's stratosphere), nor account for variable telluric transmissivity during individual integration time steps (2 seconds) of a scan, which can vary with water vapour, ice cirrus or clouds.  We therefore estimate a conservative ±20\% uncertainty on our radiometric calibration and avoid spectral regions obscured by telluric absorption features.

The scans and nods were reduced (sky subtraction, flat fielding, radiometric calibration, optical distortion corrections, interpolation over dead pixels) using the TEXES pipeline data reduction package \citep{02lacy}.  Wavelength calibration was performed by correlating telluric transmission models with the measured sky scans.  The individual target scans were shifted spectrally to correct for the Doppler shift due to the Earth-target motion and that caused by the rotation of Jupiter and Saturn.  Where longitudinal resolution was required, we worked from the individual target scans.  For the purposes of zonal mean radiances, individual scan maps were aligned spatially and co-added to create a single spectral cube.  Planetographic latitudes, System III West longitudes and emission angles were assigned to the scan maps by fitting a synthetic planetary limb to the observation, allowing interpolation and reprojection of the spectral cube onto a latitude-longitude grid. 

\begin{sidewaystable}[htdp]
\caption{Details of the Jupiter and Saturn observations used in this study.  For Saturn we performed multiple nods per setting over 70 minutes.  For Jupiter we used 2 scans per setting per night. Central longitudes are given in System III West.  The asterisk (*) indicates a medium spectral resolution observation, whereas all others were performed in low-resolution mode.}
\begin{center}
\begin{tabular}{|c|c|c|c|c|c|c|c|}
\hline
Target & Date & Setting & Time [UT] & Airmass & Humidity & Central Longitude \\
\hline
Saturn* & 2012-01-16 & 960 cm$^{-1}$ & 13:00-15:26 & 1.96-1.19  & 8-31\% & 202-285W\\
\hline
Saturn* & 2013-02-02 & 900 cm$^{-1}$ & 14:03-15:28 & 1.41-1.21 & 9\% & 80-128W \\
\hline
Saturn & 2013-02-03 & 900 cm$^{-1}$ & 13:08-14:16 &  1.71-1.35 & 24-33\% & 140W-178W\\
 &  & 960 cm$^{-1}$ & 14:27-15:42 & 1.31-1.19  & & 185W-226W \\
\hline
Saturn & 2013-02-09 & 960 cm$^{-1}$ & 15:17 & 1.20 & 20\% & 38W \\
\hline
Jupiter & 2013-02-04 & 900 cm$^{-1}$ & 06:27 &  1.02 & 24-33\% & 186W \\
 &   & 960 cm$^{-1}$ & 06:14 &  1.00 & & 179W\\
\hline
Jupiter & 2013-02-10 & 900 cm$^{-1}$ & 09:23 &  1.83 & 10-26\% & 116W \\
 &   & 960 cm$^{-1}$ & 09:16 &  1.75 & & 111W \\
 \hline
 Jupiter & 2013-02-12 & 900 cm$^{-1}$ & 09:06 & 1.73 & 15-24\% & 46W\\
 &   & 960 cm$^{-1}$ & 09:00 &  1.67 & & 42W \\
 \hline
\end{tabular}
\end{center}
\label{tab:data}
\end{sidewaystable}%

\subsection{Spatial Structure in the TEXES Observations}

\begin{figure*}[htbp]
\begin{center}
\includegraphics[width=16cm]{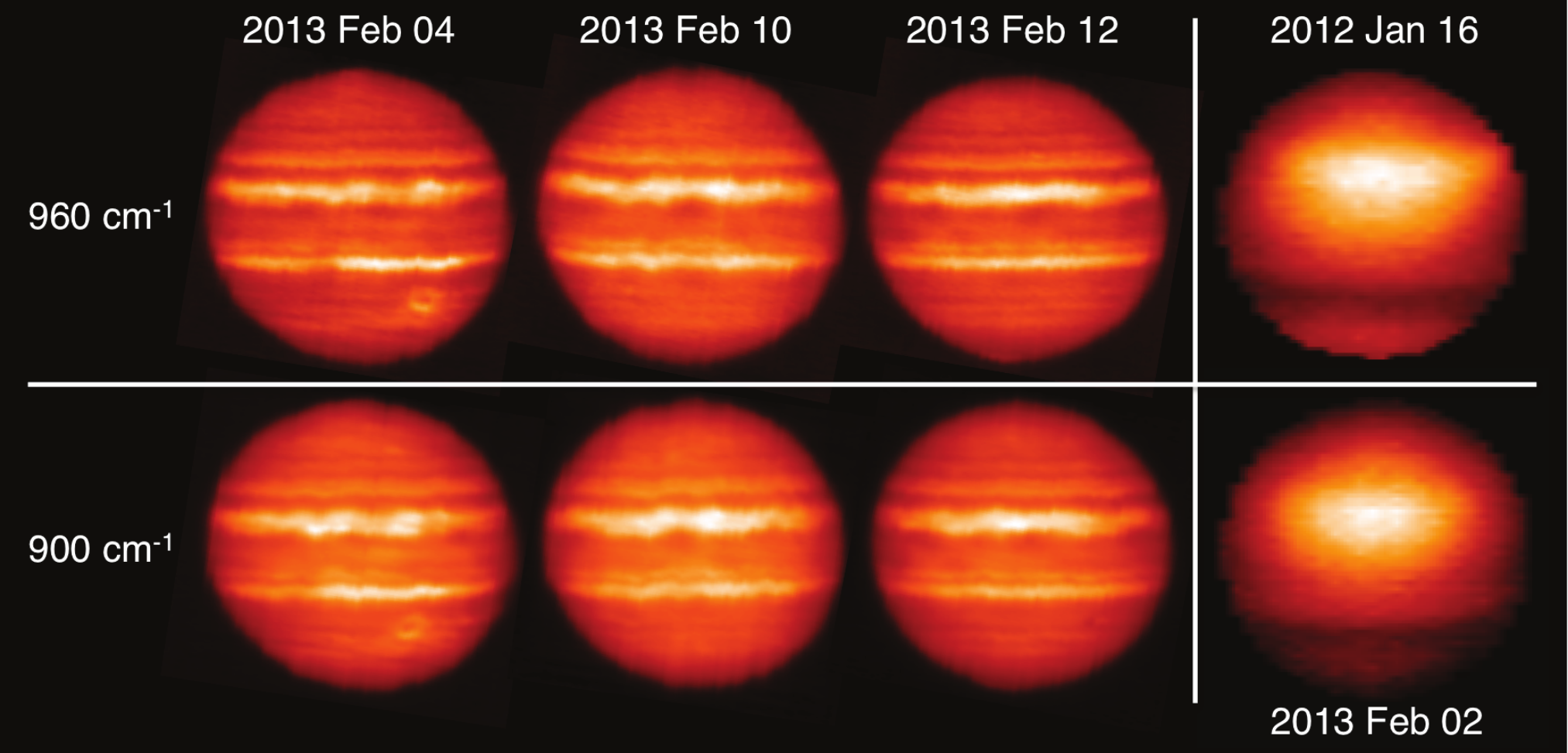}
\caption{Montage of images formed by summing radiances over the 20-\cm intervals of the two spectral settings, 900 \cm and 960 \cm, for both Jupiter and Saturn.  Spatial variations in radiance are primarily due to tropospheric temperature, but are also affected by aerosols, ammonia and phosphine.  Saturn images were constructed from medium-resolution scan maps obtained in 2012 and 2013, although the primary results of this study were obtained with low-resolution nodded observations. }
\label{images}
\end{center}
\end{figure*}

\begin{figure*}[htbp]
\begin{center}
\includegraphics[width=14cm]{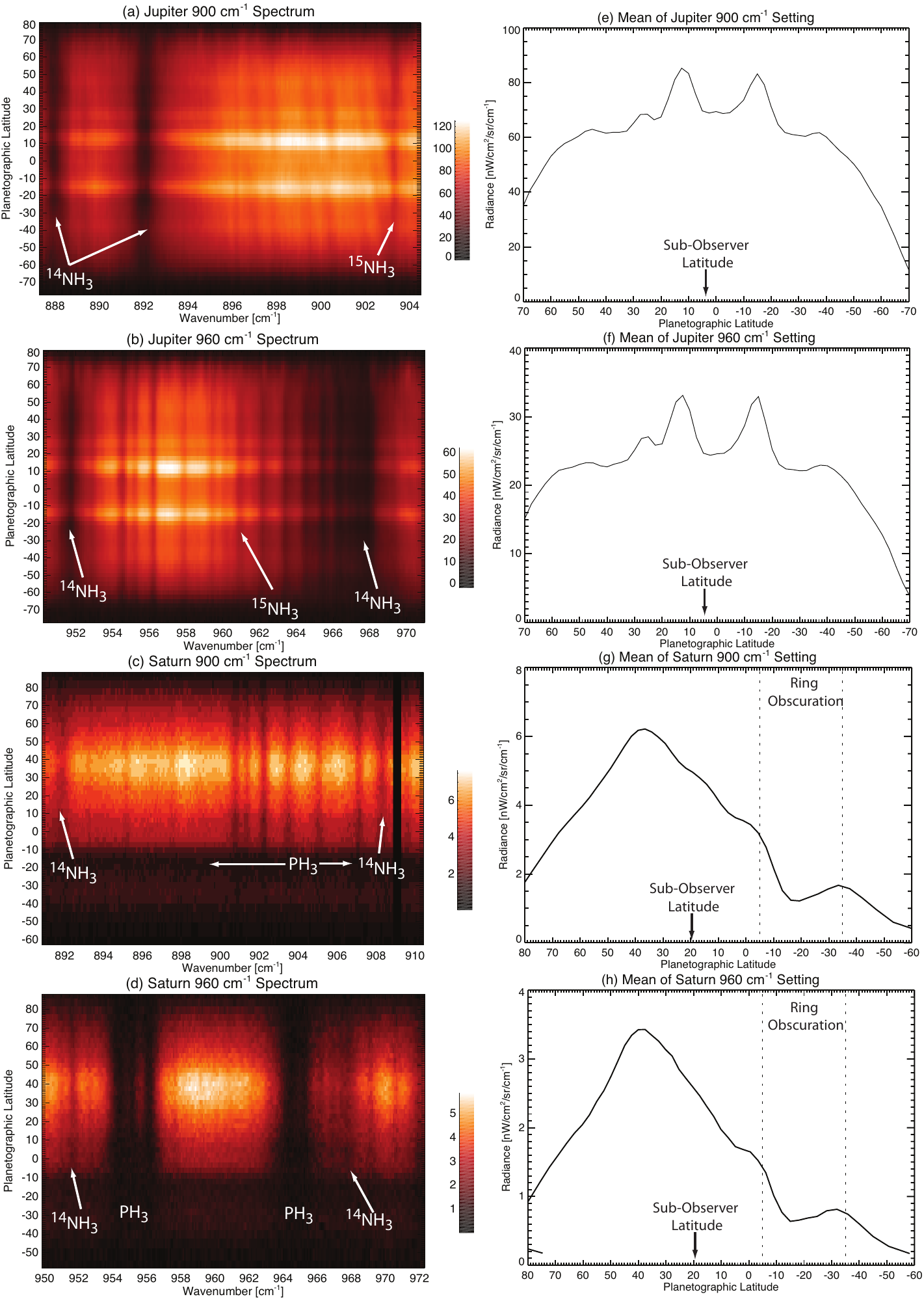}
\caption{Zonal mean radiances for Jupiter and Saturn on February 4th and 3rd, 2013, respectively.  In panels (a)-(d) the radiances are represented as spectral cubes (latitude versus wavenumber), with a scale bar indicating the calibrated radiances in nW/cm$^{2}$/sr/cm$^{-1}$ and some of the key absorption features indicated.  Panels (e)-(h) shows the zonal means averaged across the full 20-\cm band of the setting to show the primary variability (e.g., belt-zone structure on Jupiter and the storm-related peak radiance on Saturn).  The sub-observer latitude (i.e., emission angles of zero) is indicated in the right hand panels.  Diminished radiances due to Saturn's ring obscuration are indicated.  }
\label{zonalrad}
\end{center}
\end{figure*}

Figure \ref{images} provide examples of the quality of the raw spectral image cubes provided by the TEXES instrument in the 900 \cm and 960 \cm settings.  Zonal-mean radiances extracted from these cubes are then presented in Fig. \ref{zonalrad}, with absorption features of key atmospheric absorbers labelled. Telescope jitter sometimes causes offsets between adjacent scan time steps that must be corrected to reconstruct images.  The Jupiter images on all three dates show detailed banded organisation associated with the zonal wind field.  Significant longitudinal variations are seen in the bright NEB (North Equatorial Belt) associated with planetary wave activity and warm, cloud-free clearings.  The SEB (South Equatorial Belt) is now fully revived after the 2009-10 white and quiescent phase of its life cycle \citep{11fletcher_fade}, and shows both a northern and southern component, separating a cool SEB zone.  Faint equatorial warming is apparent in the otherwise cold EZ (Equatorial Zone), associated with a subtle red equatorial band observed in amateur imaging at the time.  The Great Red Spot was not observed during this sequence, but Oval BA can be seen on February 4, 2013 as a cold oval surrounded by a peripheral ring of warmer temperatures associated with atmospheric subsidence.  Polar enhancement of emission in the 960-\cm channel at the north pole on February 4 and the south pole on February 12 may be associated with ethylene emission at 950 \cm from auroral hotspots. 

TEXES spectral cubes of Saturn appear bland compared to Jupiter, both due to the smaller angular diameter but also due to the more quiescent tropospheric dynamics.  The southern hemisphere was obscured by the rings, which appear cold at these wavelengths.    When we plot the zonal mean radiance with latitude in Fig. \ref{zonalrad}, it is clear that the maximum tropospheric emission is not at the sub-observer latitude, as we might expect from limb darkening of a homogeneous planet.  Instead, the emission is localised in a band between 30-45$^\circ$N, the location of the 2010 eruption of a planet-encircling storm system \citep{11fletcher_storm}.  This region of residual tropospheric warmth may be associated with post-plume atmospheric subsidence, and appeared to encircle the whole planet with implications for tropospheric zonal wind shears \citep{12fletcher,14achterberg}.    These dynamic features on both Jupiter and Saturn are not the focus of the present study, but will be the subject of future investigations.

\subsection{TEXES Uncertainties}

\begin{figure}[htbp]
\begin{center}
\includegraphics[width=8cm]{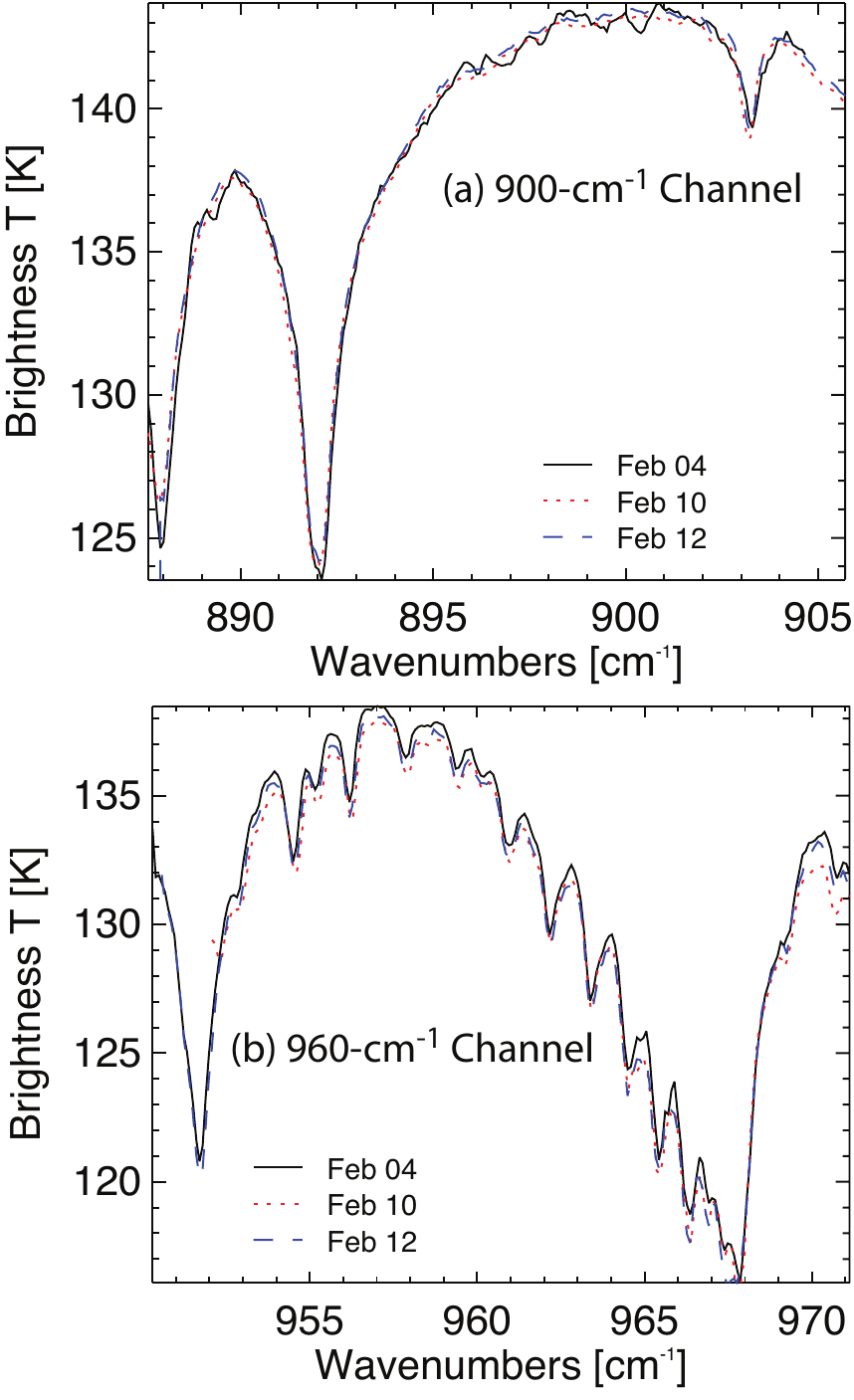}
\caption{Mean spectra of Jupiter extracted from latitudes $\pm30^\circ$ around the equator on three dates in both the 900- and 960-\cm channel.  Radiances were converted to brightness temperatures to demonstrate consistency from night to night of better than 1 K.}
\label{compspx}
\end{center}
\end{figure}

Ground-based constraints on the $^{15}$N/$^{14}$N ratio require a careful analysis of uncertainties associated with the TEXES spectra.  Modelling uncertainties will be considered in Section \ref{model}.  For Jupiter, we constructed an average spectrum by co-adding data between $\pm30^\circ$ latitude and within $\pm10^\circ$ of the central longitude (i.e., a low-latitude average of the spectra shown in Fig. \ref{zonalrad}).  For Saturn we used central-meridian data from the nodded spectra between latitudes of 10$^\circ$N and 60$^\circ$N.  Mean brightness-temperature spectra from the three epochs of Jupiter observations are shown in Fig. \ref{compspx}, demonstrating that independent observations are radiometrically calibrated to precisions better than 1 K.  However, precision does not imply accuracy, as telluric transmission is not fully accounted for using the flat fielding and absolute calibration method described by \citet{02lacy}.

\begin{figure}[htbp]
\begin{center}
\includegraphics[width=8cm]{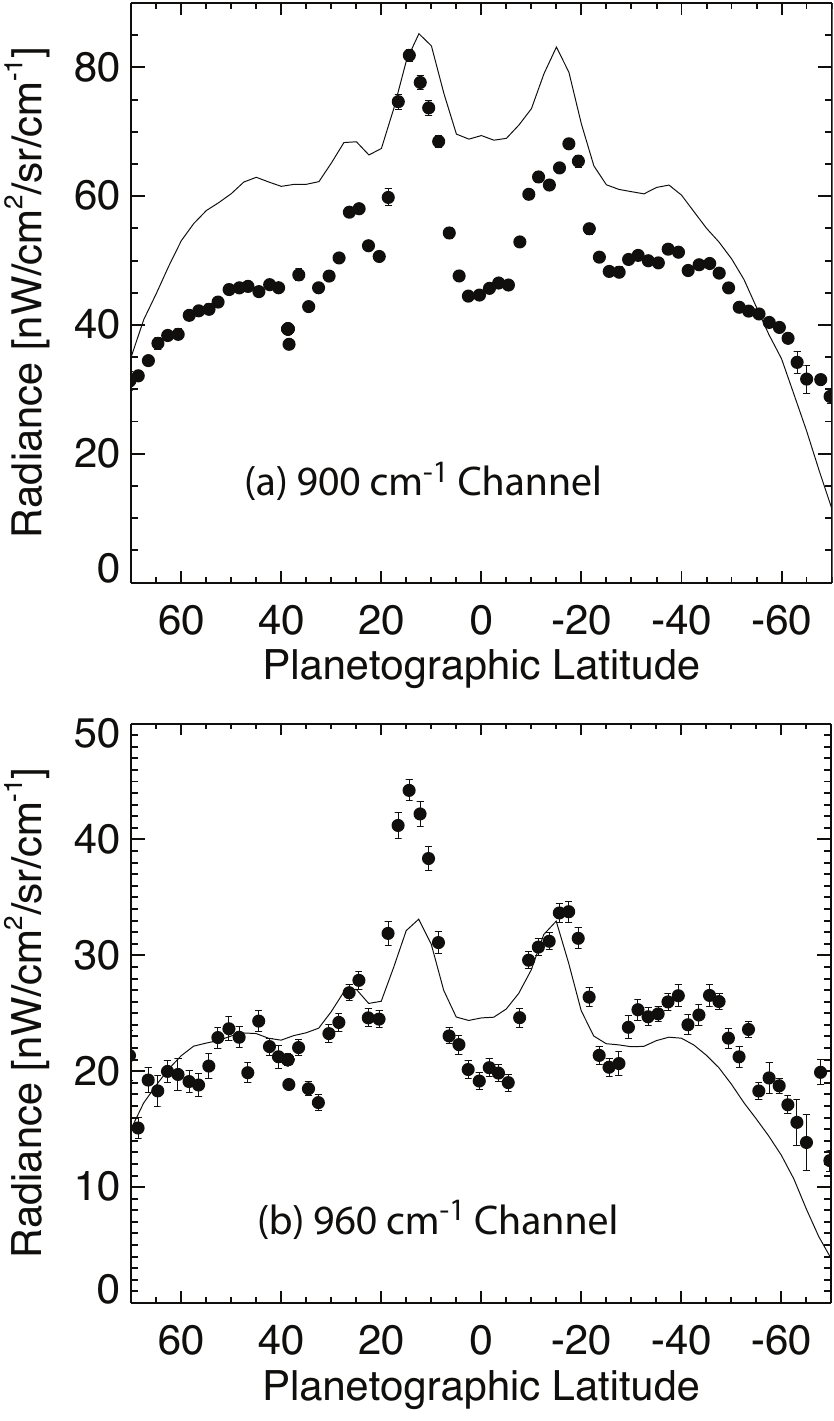}
\caption{Comparison of TEXES Jupiter zonal mean radiances (solid line) at 900 and 960 \cm spectral settings with Cassini/CIRS radiances obtained in 2000 (circles with error bars).  The 900-\cm observations from TEXES need to be scaled by a factor of 0.8 to match the CIRS radiances, whereas the 960-\cm setting needs no such adjustment. }
\label{cirscomp}
\end{center}
\end{figure}

The zonal mean radiances for Jupiter in Fig. \ref{zonalrad} are compared to Cassini/CIRS observations from the 2000 flyby in Fig. \ref{cirscomp}, averaging the CIRS spectra over the same spectral range.  The diminution of radiance towards the poles is an artefact of the strong limb darkening observed by both CIRS and TEXES.  The zonal banding matches qualitatively, but we found that the 900-\cm setting had to be scaled by a factor of 0.8 to match the CIRS radiances.  As global-scale temperature changes between 2000 and 2013 are unlikely, this suggests that our spectrophotometric calibration, although precise and reproducible, is uncertain by factors of $\pm20$\%.  Conversely the 960-\cm setting required no such scaling, possibly because the 960-\cm setting is near the centre of the filter bandpass, whereas the 900-\cm setting is near the edge.  Unfortunately the CIRS Saturn spectra averaged over 2012-2013 have insufficient signal-to-noise in this spectral region to permit a similar comparison, so we again assume a $\pm20$\% uncertainty for the Saturn data.  As the 900-\cm scaling was the same for each night, the cause is likely to be instrumental transmissivity effects unaccounted for in the reduction pipeline, rather than random atmospheric variations.  This systematic uncertainty has severe consequences for our ability to derive absolute abundances from the TEXES spectra, as we shall see in Section \ref{results}.

In addition to systematic offsets, TEXES random uncertainties were estimated using four different methods.  Firstly, we compared the simple standard deviations on the mean spectra with a brute-force method of shifting the spectrum by one spectral pixel and differencing.  Both techniques result in random uncertainties of 3\% in clear atmosphere, growing to arbitrarily large values near telluric features.  Second, we used the individual Jupiter scan maps and compared spectra in the sky emission background (averages over 25 pixels in each of the four corners of the images in Fig. \ref{images}, and found random uncertainties of 1.0-1.5\% for the 900 \cm setting and 4-5\% for the 960 \cm setting.  Third, we used telluric transmission $\tau$ calculated via the ATRAN model \citep{92lord} and the empirical formula of \citet{05greathouse} for the uncertainty, $\sigma=(\sqrt{1.1-\tau})/\tau$, where the factor of 1.1 accounts for the emissivity of the instrument window and telescope.  This resulted in more conservative clear-sky uncertainties around 5\%.   Fourth, we used the measured sky emission from a small section of the cube centred on the sub-observer point, $\tau_m$, and the formula $\sigma_m=(\sqrt{1.0-\tau_m})/\tau_m$, where the 1.0 factor is smaller than before because the measurements already include flux losses due to instrumental effects.  This final method produced the most conservative uncertainties of up to 10\% in  clear sky, but also includes the systematic changes due to the instrument effects.  In Section \ref{results} we will present our results both for method one (taking the 3\% standard deviation on the mean spectrum, increasing to arbitrarily large values in regions of telluric absorption) and three (uncertainty based on the calculated transmission).

\section{Spectral Modelling}
\label{model}

Atmospheric parameters were derived from the TEXES spectral cubes using a radiative-transfer and spectral-retrieval algorithm \citep[NEMESIS,][]{08irwin}.  This software performs a spectral inversion via the optimal-estimation approach \citep{00rodgers}, minimising a two-term cost function comprised of the residual fit to the data and our \textit{prior} knowledge of the atmospheric state vector.  The latter constraint ensures smooth and physically realistic atmospheric structures.  Furthermore, by inflating the measurement uncertainties in regions of low telluric transmission (see Section \ref{data}) we ensure that the retrievals are weighted towards clear-sky spectral windows and ignore those corrupted by terrestrial contamination.  Reference \textit{a priori} temperature and composition profiles have been previously described in detail for Jupiter \citep{09fletcher_ph3,11fletcher_trecs} and Saturn \citep{10fletcher_seasons,12fletcher}.  In particular, the reference $T(p)$ were defined on 120 pressure levels equally spaced in $\log(p)$ between 1 $\mu$bar and 10 bar, and are based on a low-latitude average of Cassini/CIRS temperature results.  Details of the vertical profiles of ammonia and phosphine are introduced below.

Sources of spectral line data are identical to those presented in Table 4 of \citet{12fletcher}, and were used to pre-tabulate $k$-distributions (ranking absorption coefficients according to their frequency distributions) specific to the two low-resolution TEXES spectral settings (900 \cm and 960 \cm) using the direct sorting method of \citet{89goody_ck}.  These $k$-distributions were convolved with instrument functions to represent the specific characteristics of TEXES.  However, the TEXES instrument function is not precisely constrained:  diffraction convolved with a box car of the slit width should provide approximately Lorentzian wings on the lines, but optical aberrations might also contribute and we assume the true instrument function to lie somewhere between a Gaussian and a Lorentzian.  Furthermore, the spectral resolution provided by the grating equation in Section \ref{data} is an approximation.  We therefore conducted preliminary retrieval comparisons for the Jupiter and Saturn spectra with Gaussian-, Lorentzian- and triangle-convolved $k$-distributions with a variety of full-width at half maxima (FWHM) between 0.20 and 0.40 \cm for both channels.  However, we found negligible differences in the goodness-of-fit ($\chi^2$) for these three different instrument functions, and the best-fitting FWHM differed substantially between the four different dates.  The optimal FWHM were apparently driven by random noise on each date.  As an alternative, we attempted to convolve synthetic terrestrial ATRAN spectral models (with Lorentzian line shapes and a 0.10-\cm width) with Gaussians of variable width until the model reproduced the measured sky lines.  However, our spectral channels are chosen so that the sky lines are very weak, so this technique only worked for the 960-\cm channel (converging on a resolution of 0.32 \cm) and not for the 900-\cm channel.  Given the difficulty in extracting the resolution from the measurements, our solution was to use a Gaussian instrument function and the FWHM predicted by the grating equation (0.31 \cm and 0.36 \cm for the 900- and 960-\cm channels, respectively).  

In this preliminary phase of TEXES modelling we discovered that the manual identification of sky lines, coupled with corrections for Doppler shifting by the reduction pipeline, did not provide the necessary wavelength accuracy for our purposes.  For the mean low-latitude spectra for each channel and date, we varied the wavelength calibration using a shift-and-stretch approach to minimise the $\chi^2$.  The required shifts were $<0.01$ \cm and $<0.02$ \cm for the 900- and 960-\cm channels, respectively, but were required to provide accurate fits to the data.  These shifts were applied to generate wavelength-corrected mean and zonally-averaged spectra for subsequent analysis.

\begin{figure*}[htbp]
\begin{center}
\includegraphics[width=16cm]{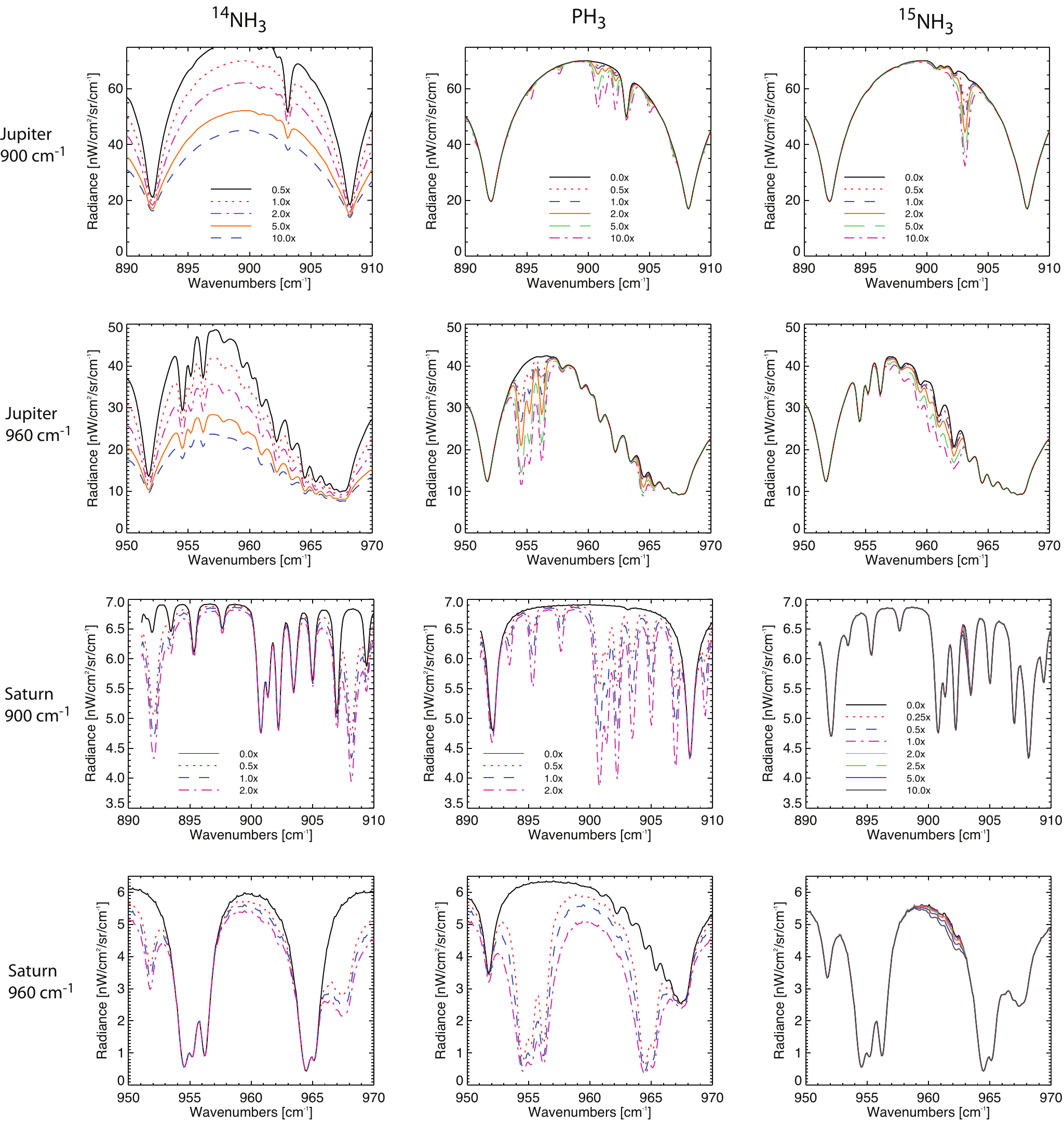}
\caption{Forward-modelled Jupiter (top panels) and Saturn (bottom panels) spectra in the 900- and 960-\cm spectral settings, showing the effects of scaling the reference $^{15}$NH$_3$ (right column), $^{14}$NH$_3$ (left column) and PH$_3$ (central column) profiles.  A key for the Jupiter scale factors can be found in the top row, a key for the Saturn scale factors can be found in the third row.  This highlights the difficulty in obtaining the Saturnian value, although enhanced $^{15}$NH$_3$ causes peak erosion near 903 \cm and spectral features near 960 \cm which will be key to the upper limits determined in this study.  These spectra use a low-latitude average temperature and composition and nadir viewing geometry. }
\label{fmodel}
\end{center}
\end{figure*}

\begin{figure*}[htbp]
\begin{center}
\includegraphics[width=16cm]{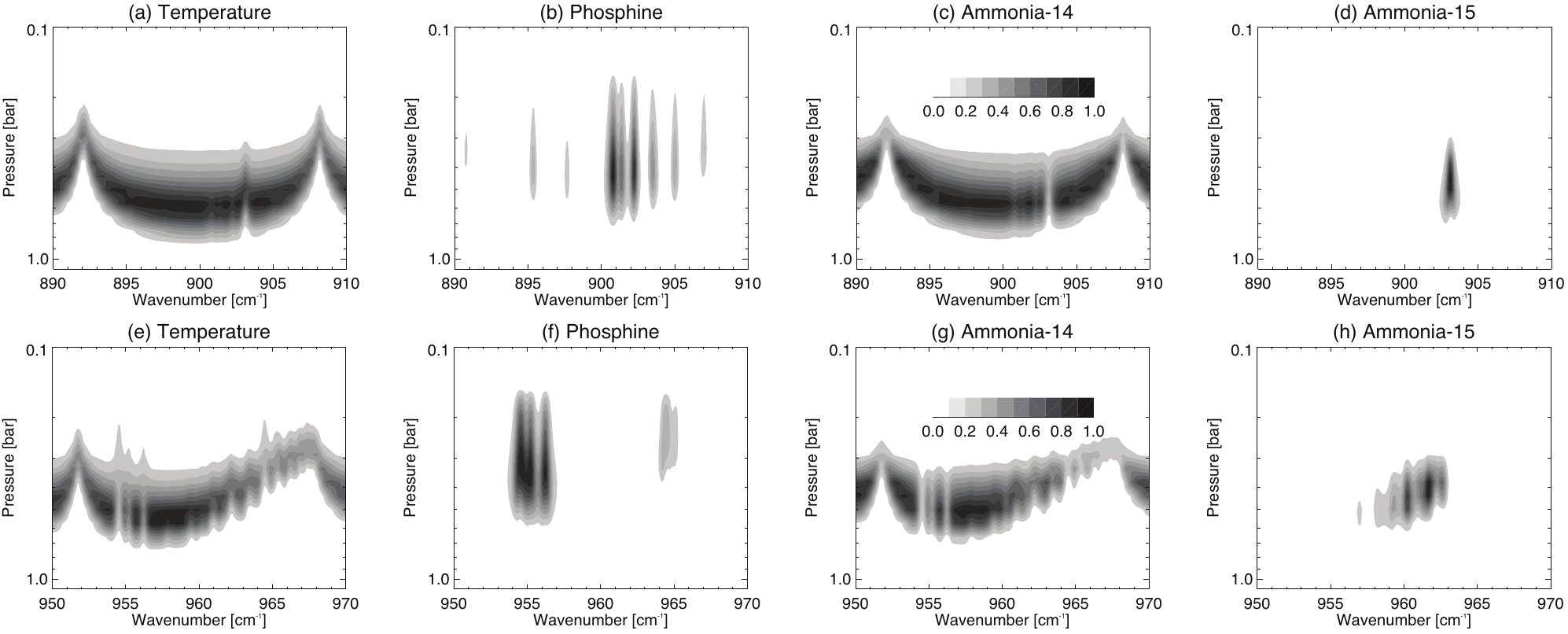}
\caption{Jacobians, or functional derivatives, showing the sensitivity of the TEXES Jupiter spectra to temperature variations and abundance variations for each of the gases ($^{15}$NH$_3$, $^{14}$NH$_3$ and PH$_3$).  The 900-\cm channel is given on the top row; the 960-\cm channel on the bottom row. The jacobians in each panel have been normalised within the spectral range, so the key in panels (c) and (g) is representative of all eight panels.}
\label{jupcontrib}
\end{center}
\end{figure*}

\begin{figure*}[htbp]
\begin{center}
\includegraphics[width=16cm]{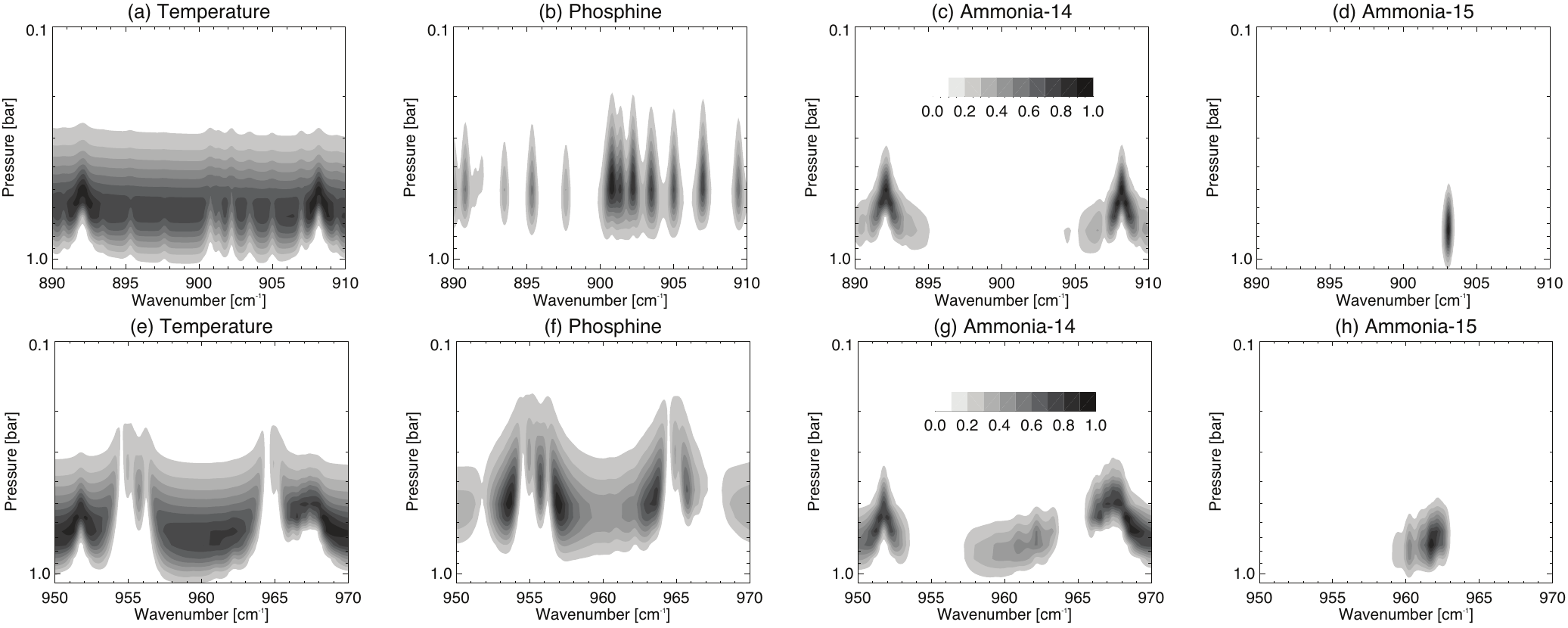}
\caption{Jacobians, or functional derivatives, showing the sensitivity of the TEXES Saturn spectra to temperature variations and abundance variations for each of the gases ($^{15}$NH$_3$, $^{14}$NH$_3$ and PH$_3$).   The 900-\cm channel is given on the top row; the 960-\cm channel on the bottom row.  The jacobians in each panel have been normalised within the spectral range, so the key in panels (c) and (g) is representative of all eight panels.}
\label{satcontrib}
\end{center}
\end{figure*}

Forward-modelled spectra in Fig. \ref{fmodel} indicate the locations of the ammonia and phosphine absorption features in the 900- and 960-\cm channels.  We also show the effects of varying the $^{15}$N/$^{14}$N ratio from zero to a factor of $10\times10^{-3}$ (encompassing the Jupiter-like $2\times10^{-3}$ and Titan-like $6.0\times10^{-3}$ values).  In the 900-\cm channel $^{14}$NH$_3$ comes from lines at 892 and 908 \cm, with $^{15}$NH$_3$ perturbing the spectra near 903 \cm.  At 960 \cm the $^{14}$NH$_3$ is from 952 \cm and longward of 966 \cm, with $^{15}$NH$_3$ affecting the spectrum between 959 and 963 cm$^{-1}$.  Fig. \ref{fmodel} confirms the small but detectable effects of $^{15}$NH$_3$ on Saturn's phosphine-dominated spectrum (e.g., peak erosion at 903 \cm and spectral features at 960 cm$^{-1}$).  Figs. \ref{jupcontrib}-\ref{satcontrib} demonstrate the altitude sensitivity of the TEXES spectra by presenting jacobians, or functional derivatives $dR/dx$ (where $R$ is the spectral radiance and $x$ is the atmospheric state vector, representing either temperature or a mole fraction profile).  This highlights the overlapping contributions of temperature (which covers most of the range) with the molecular signatures, and difficulties in disentangling temperature and composition will be discussed in Section \ref{results}.  Fig. \ref{jupcontrib} confirms the findings of \citet{04fouchet} for Jupiter, who demonstrated that the different isotopologues sense similar pressures in the 400-600-mbar range.  The Jupiter jacobians were calculated using an aerosol layer near 800 mbar based on Cassini/CIRS retrievals \citep{09fletcher_ph3}.  Saturn's continuum radiance in Fig. \ref{satcontrib} \citep[computed for aerosol-free conditions following ][]{09fletcher_ph3} originates in the 500-800 mbar range, with radiance in the PH$_3$ absorption cores penetrating up towards 200 mbar.  Saturn's $^{15}$NH$_3$ signatures are formed in the 600-900 mbar range, slightly deeper but overlapping with the $^{14}$NH$_3$ signatures in the 400-800 mbar range (deepest in the line wings).    In both the Jupiter and Saturn cases, the temperature and gaseous jacobians are sufficiently distinct to permit separation of these parameters in the retrievals, as presented in Section \ref{results}.

\section{Results}
\label{results}

Measurements of temperature and atmospheric composition from the TEXES data are complicated by (i) the absolute radiometric uncertainty of $\pm20$\% on each channel; (ii) the use of narrow 20-\cm windows leading to degeneracies between tropospheric temperature, aerosol opacity and the molecular abundances; and (iii) the low sensitivity to the vertical profiles of PH$_3$ and NH$_3$.  In the sections that follow, we conduct extensive tests to understand the sensitivity of our results to these complications.  For both Jupiter and Saturn, we start with simultaneous fitting of the 900- and 960-\cm channels to obtain a `global solution' (Section \ref{jup_glob} and \ref{satglob}), before honing the spectral fit in the vicinity of $^{15}$NH$_3$ lines and fitting to determine the $^{15}$N/$^{14}$N ratio (Sections \ref{jup_15N} and \ref{sat_15N}).  Finally, we repeat the fitting for latitudinally-resolved spectra to understand the spatial variability in Section \ref{zonal}.

\subsection{Global Solution for Jupiter}
\label{jup_glob}
An ideal retrieval would allow us to independently determine tropospheric temperatures, aerosol opacity and parameterised vertical distributions of phosphine and ammonia \citep[parameterised as a deep mole fraction $q_0$ at $p>p_0$ and a fractional scale height $f$ at $p<p_0$,][]{07fletcher_ph3}.  The transition pressures $p_0$ for ammonia and phosphine are poorly known, so these parameters were varied on a two-dimensional grid from 300 to 1000 mbar during retrievals of temperature, $q_0$ and $f$ for PH$_3$ and NH$_3$.  We scaled the total opacity of a compact aerosol layer of NH$_3$ ice (10-$\mu$m radius), based at 800 mbar.  We need to allow the temperature to vary in order to achieve an adequate reproduction of the measurements, despite the lack of independent tropospheric temperature constraints from the collision-induced H$_2$-He continuum at $\lambda>16 \mu$m.  Furthermore, the retrieved departure of the TEXES $T(p)$ from that derived by CIRS would prove essential in narrowing the broad solution space due to the TEXES radiometric uncertainties (see below).   

The jovian phosphine lines could be adequately reproduced with $p_0$ at any altitude - the deeper the transition, the larger the permitted values of $q_0$ and $f$, and these changes were so complementary that no information on $p_0$ could be obtained.  Instead, we reduced the under-constrained problem to a single parameter - scaling a photochemical PH$_3$ profile from J. Moses (\textit{personal communication}) that was based on the CIRS-derived PH$_3$ abundance of \citet{09fletcher_ph3}.  These chemical profiles have the added benefit of a smooth and gradual transition from the well-mixed zone to the region of photochemical destruction, rather than a sharp change at $p_0$.

\begin{figure*}[htbp]
\begin{center}
\includegraphics[width=16cm]{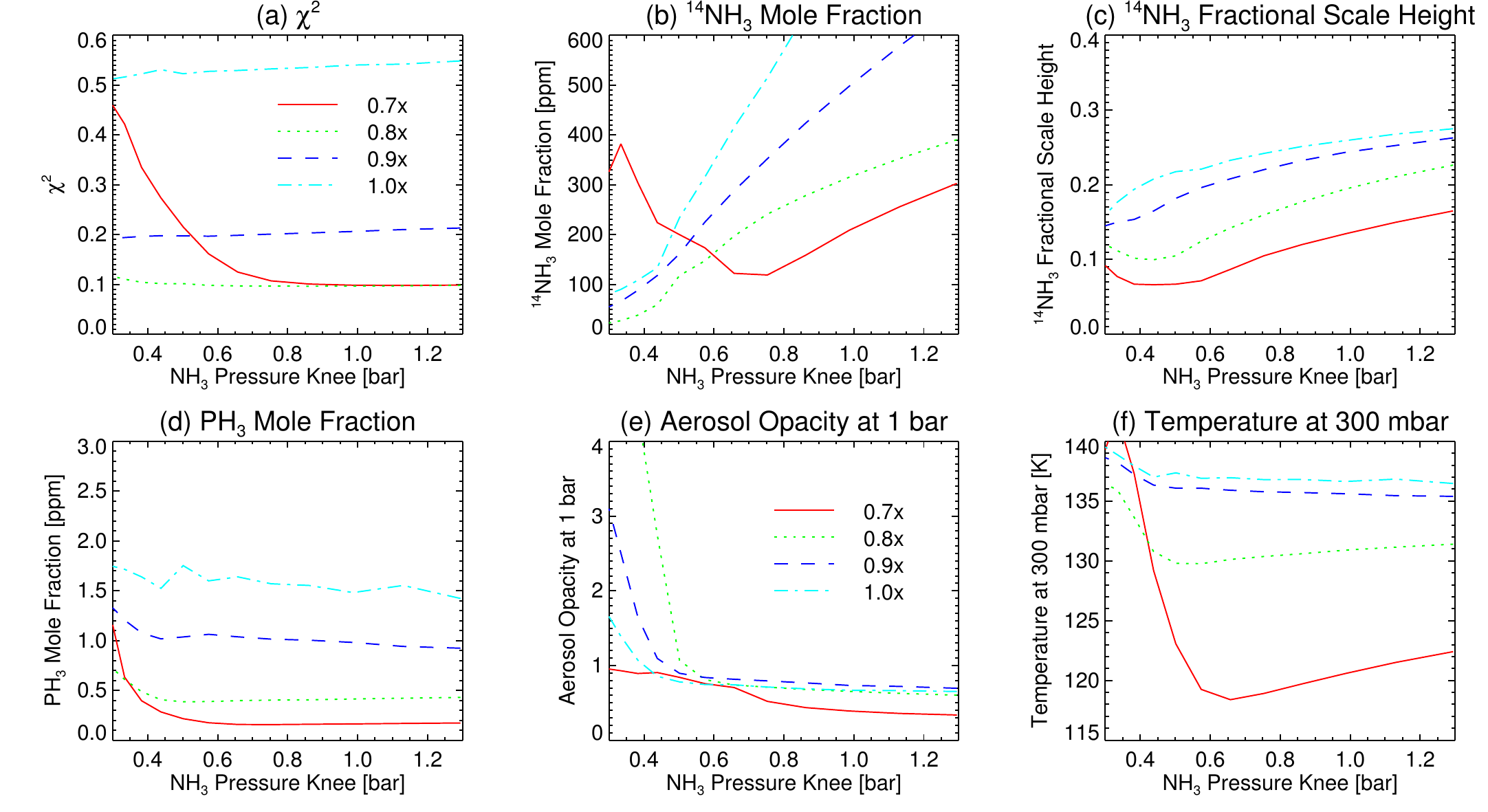}
\caption{The behaviour of retrieved parameters and goodness-of-fit $\chi^2$ for different assumptions about the transition pressure for Jupiter's NH$_3$ vertical distribution.  Each plot features four curves, representing different results for the radiometric scalings of the 900-\cm channel by factors of 0.7-1.0.  The radiance in the 960-\cm channel was unaltered.  The $\chi^2$ values favour the radiometric scalings of 0.7-0.8, and these constrain $p_0>800$ mbar.  Results are plotted for February 4th 2013, but similar behaviour was seen on all three dates.}
\label{nh3knee}
\end{center}
\end{figure*}

Given the dominance of ammonia in Jupiter's mid-IR spectrum, it might be expected that the TEXES retrievals could constrain its transition pressure, $p_0$.  We performed simultaneous retrievals of $T(p)$, scale factors for the aerosol opacity and PH$_3$ distribution, and $q_0$ and $f$ for NH$_3$, for a variety of transition pressures from 300-1200 mbar.  Given the radiometric uncertainty, we repeated this test for all three dates with radiometric scale factors of 0.7-1.2 applied to each of the 900- and 960-\cm channels (i.e., 25 scale factor combinations, 10 values of $p_0$ and 3 separate Jupiter observation dates).  The behaviour of $\chi^2$ and the retrieved parameters for a subset of these results are shown in Fig. \ref{nh3knee}.  In this instance, we fixed the calibration of the 960-\cm channel and only show results for scaling the 900-\cm channel by factors between 0.7 and 1.0.  Best fits are obtained when scale factors of 0.7-0.8 are used for the 900-\cm channel (consistent with the necessity for scaling the TEXES data to match the CIRS observations in Section \ref{data}).  The best-fitting $p_0$ is extremely sensitive to the radiometric scaling, and values in the expected range \citep[i.e., close to the NH$_3$ condensation altitude of 840 mbar for a $3\times$ enrichment in nitrogen over solar,][]{99atreya} are only obtained for scalings of 0.7-0.8.  The retrieved $q_0$ and $f$ generally increase as the transition pressure moves deeper (Fig. \ref{nh3knee}b-c), while the PH$_3$ and aerosol scale factors are largely insensitive to the NH$_3$ $p_0$ for pressures exceeding 500 mbar (Fig. \ref{nh3knee}d-e).  For very shallow transition pressures (300-500 mbar), the aerosol opacity, phosphine abundance and temperature must all increase to compensate for poor fitting of the NH$_3$ lines.  Given the uncertainty on the radiometric scaling, we cannot independently constrain $p_0$ for NH$_3$, so we fixed the ammonia $p_0=800$ mbar, consistent with the expected cloud condensation altitude \citep{99atreya, 98banfield, 05matcheva, 04wong}.  

\begin{figure*}[htbp]
\begin{center}
\includegraphics[width=16cm]{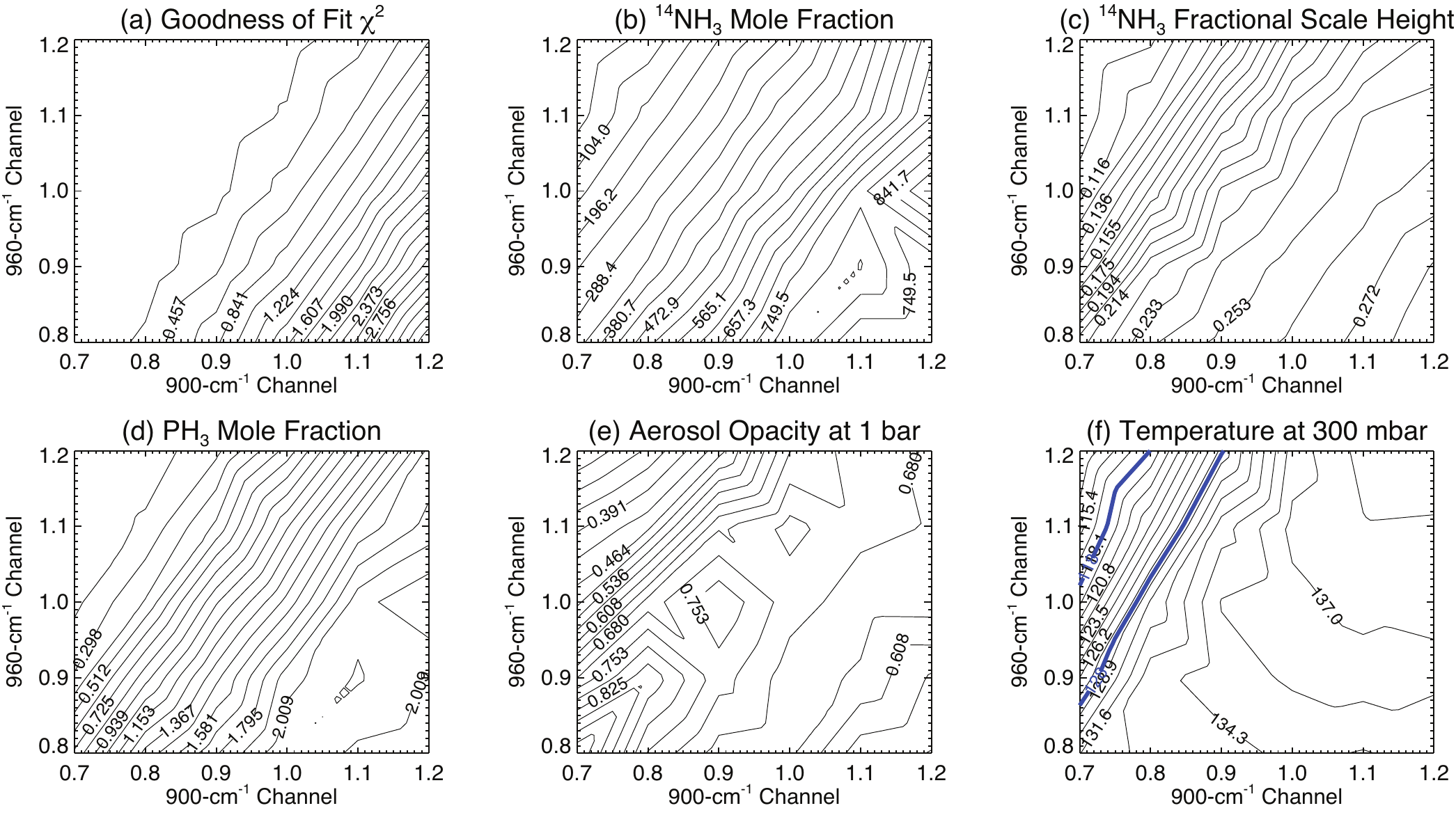}
\caption{Contours highlighting the effect of $\pm20$\% radiometric uncertainties (i.e., scale factors applied to the calibration of the 900- and 960-\cm channels) on the retrieved parameters (temperature, scale factors for aerosol opacity and phosphine, parameterised ammonia).  Also shown as blue lines is the more limited parameter space defined by a 5-K deviation from Cassini/CIRS temperature measurements at 300 mbar (between the thick dashed contour lines in panel (f)), showing that only a subset of the parameter space provides consistency with Cassini measurements.  }
\label{radscale}
\end{center}
\end{figure*}

\begin{figure}[htbp]
\begin{center}
\includegraphics[width=8cm]{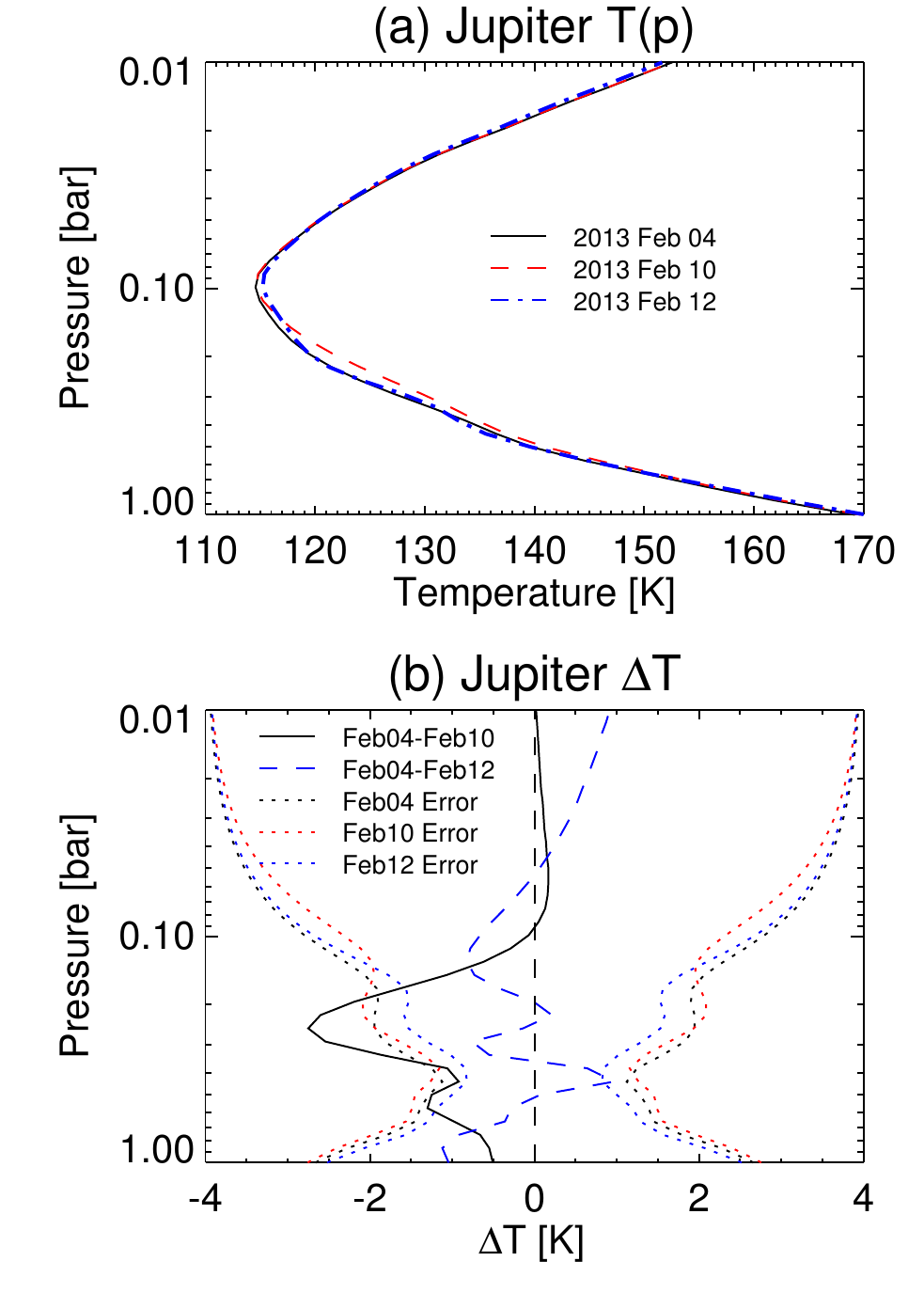}
\caption{Panel (a) shows the consistency of $T(p)$ profiles over three different observing dates for Jupiter, scaling the 900-\cm radiometric calibration by a factor of 0.8 and leaving the 960-\cm channel unaltered.  Temperature differences are less than 3 K in panel (b), a similar order-of-magnitude to the retrieval uncertainty (shown as dotted lines). }
\label{Tprofile}
\end{center}
\end{figure}

\begin{figure}[htbp]
\begin{center}
\includegraphics[width=8cm]{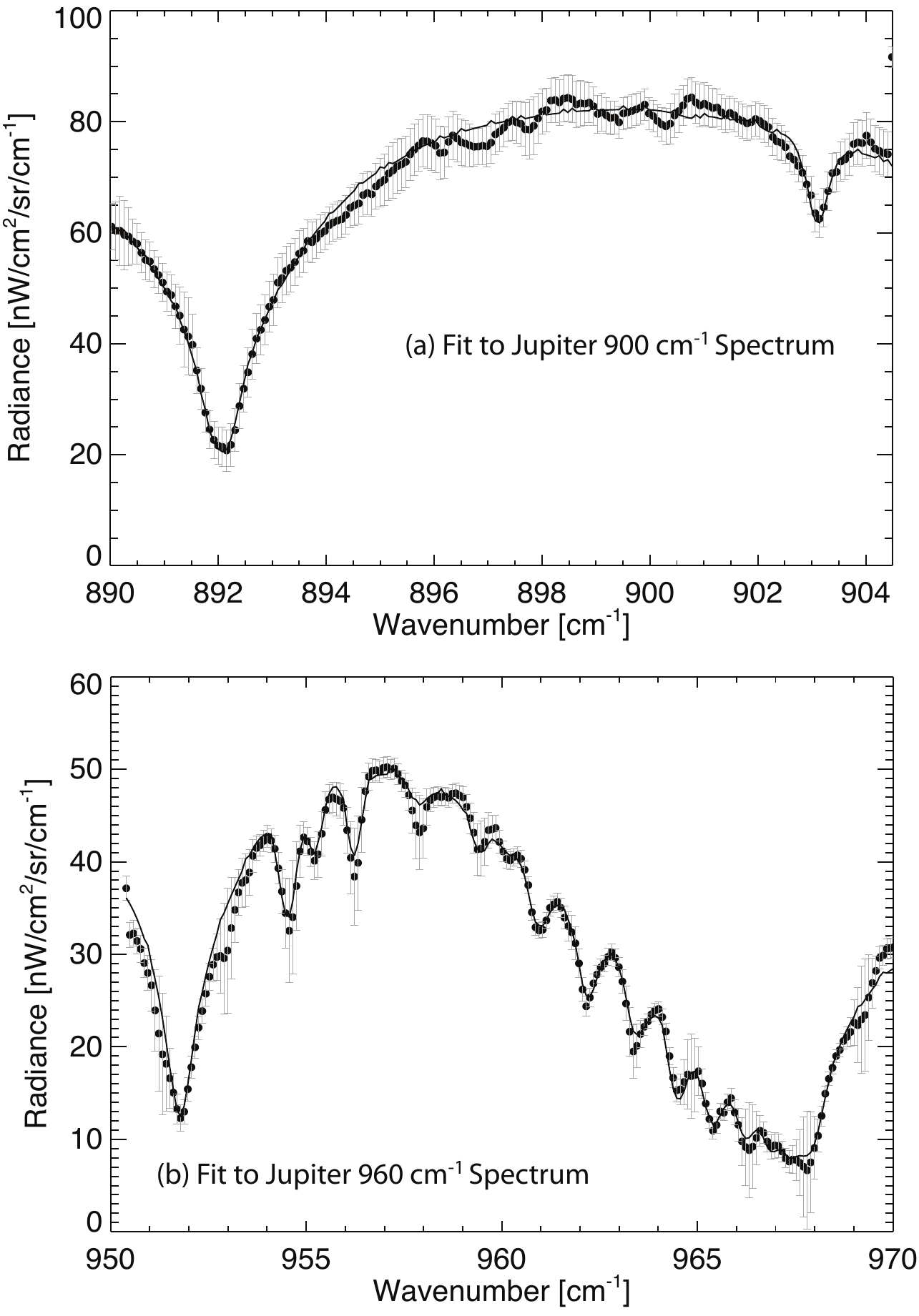}
\caption{Comparing the TEXES measurements (circles with error bars) and the best spectral fits (solid line) to the two channels.  The 900-\cm channel has been scaled by a factor of 0.8.  Error bars are inflated in regions of telluric contamination to ensure that the retrievals are weighted towards clear-sky regions of the spectrum.  Measurements have been averaged over $\pm30^\circ$ latitude and models assume a single representative emission angle of $17^\circ$.}
\label{jup_spxfit}
\end{center}
\end{figure}

The retrieved temperatures in Fig. \ref{nh3knee} show a substantial sensitivity to the radiometric scaling (14-K differences between the 0.7 and 1.0 scale factors for the 900-\cm channel).  Indeed, these differences in $T(p)$ result in order-of-magnitude changes in the retrieved ammonia abundance.  By itself, this implies that TEXES data in these two channels cannot provide robust temperature and composition retrievals.  However, globally-averaged tropospheric temperatures are not expected to deviate substantially from those measured by Cassini or Voyager, so we use this assumption to hone in on an acceptable range of radiometric scalings.  Fig. \ref{radscale} shows contours of $\chi^2$, ammonia $q_0$ and $f$, scale factors for aerosol opacity and PH$_3$ and temperatures at 300 mbar when we rerun our grid of radiometric scalings.  The region of parameter space consistent with CIRS $T(p)$ measurements $\pm5$ K is also indicated, confirming that the 900-\cm channel must be reduced by factors of 0.7 to 0.8 relative to the TEXES calibration to achieve sensible results.  All three dates showed the same trend, with the best-fitting $T(p)$ shown in Fig. \ref{Tprofile} corresponding to the spectral fits in Fig. \ref{jup_spxfit} and the compositions in Table \ref{jupcomp}.  Uncertainties shown in this table are due to random measurement error, and do not account for systematic radiometric uncertainty.  They highlight the difficulty in constraining aerosol optical depth (the uncertainty is a comparable size to the central value) and phosphine (which varies by a factor of two between the observations).  Given that the spatially-averaged Jupiter spectra in Fig. \ref{compspx} showed some small differences from night to night, longitudinal variability could be the source of these changes.

\begin{sidewaystable}[htp]  
\caption{Best-fitting compositional results derived from simultaneous fits to the 900- and 960-\cm channels for Jupiter on three dates.  Results are quoted for the case where data in the 900-\cm channel were scaled by a factor of 0.8.}
\begin{center}
\begin{tabular}{c|c c c }
\hline
 & Feb 4 & Feb 10 & Feb 12\\
\hline
NH$_3$ $q_0$ at $p>800$ mbar [ppm]& 	257$\pm$112	& 270$\pm$106 &	271$\pm$106 \\
NH$_3$ Fractional Scale Height $f$ &	0.167$\pm$0.018	&0.227$\pm$0.017	&0.190$\pm$0.021 \\
PH$_3$ at 500 mbar [ppm] &	0.41$\pm$0.09	&0.83$\pm$0.12	&0.58$\pm$0.14 \\
Aerosol Optical Depth at 1 bar &	0.70$\pm$0.63&	1.0$\pm$0.90&	0.81$\pm$0.76 \\
 \hline
\end{tabular}
\end{center}
\label{jupcomp}
\end{sidewaystable}%

\subsection{Global Solution for Saturn}
\label{satglob}
Fits to the averaged low-resolution Saturn spectra on February 3, 2013 followed similar techniques to those for Jupiter.    However, unlike Jupiter, where we had three observing dates to compare and a consistency check with Cassini/CIRS measurements, the Saturn spectra were analysed in isolation.  As before, we assumed a random uncertainty on the spectrum of 3\% in clear-sky regions (arbitrarily large in regions affected by telluric contamination), and repeated the retrievals for radiometric scalings of $\pm20$\% for each of the 900- and 960-\cm channels.  The Saturn spectra have been averaged around the sub-observer point, between $10^\circ$N and $60^\circ$N.  Tropospheric temperatures and parameterised profiles of ammonia and phosphine were retrieved simultaneously for each radiometric scaling (i.e., 25 combinations).  Although good fits could be obtained for every combination, there was a general preference for the 900-\cm channel to be decreased in radiance compared to the 960-\cm channel, consistent with the results from the Jupiter modelling.  If we assume that the radiometric calibration of the 960-\cm channel is accurate, then best fits were obtained with the 900-\cm channel shifted down by 20-30\%.  We conclude that systematic shifts in the absolute calibration are affecting the Jupiter and Saturn spectra in the same way.  However, there are substantial differences in absolute temperatures and gaseous abundances derived from the TEXES spectra for different radiometric scalings.  To get around this problem, the observations were designed so that both $^{14}$NH$_3$ and $^{15}$NH$_3$ were captured in the same setting, so the ratio should be unaltered even if the absolute quantities differ.  

\begin{figure*}[htbp]
\begin{center}
\includegraphics[width=16cm]{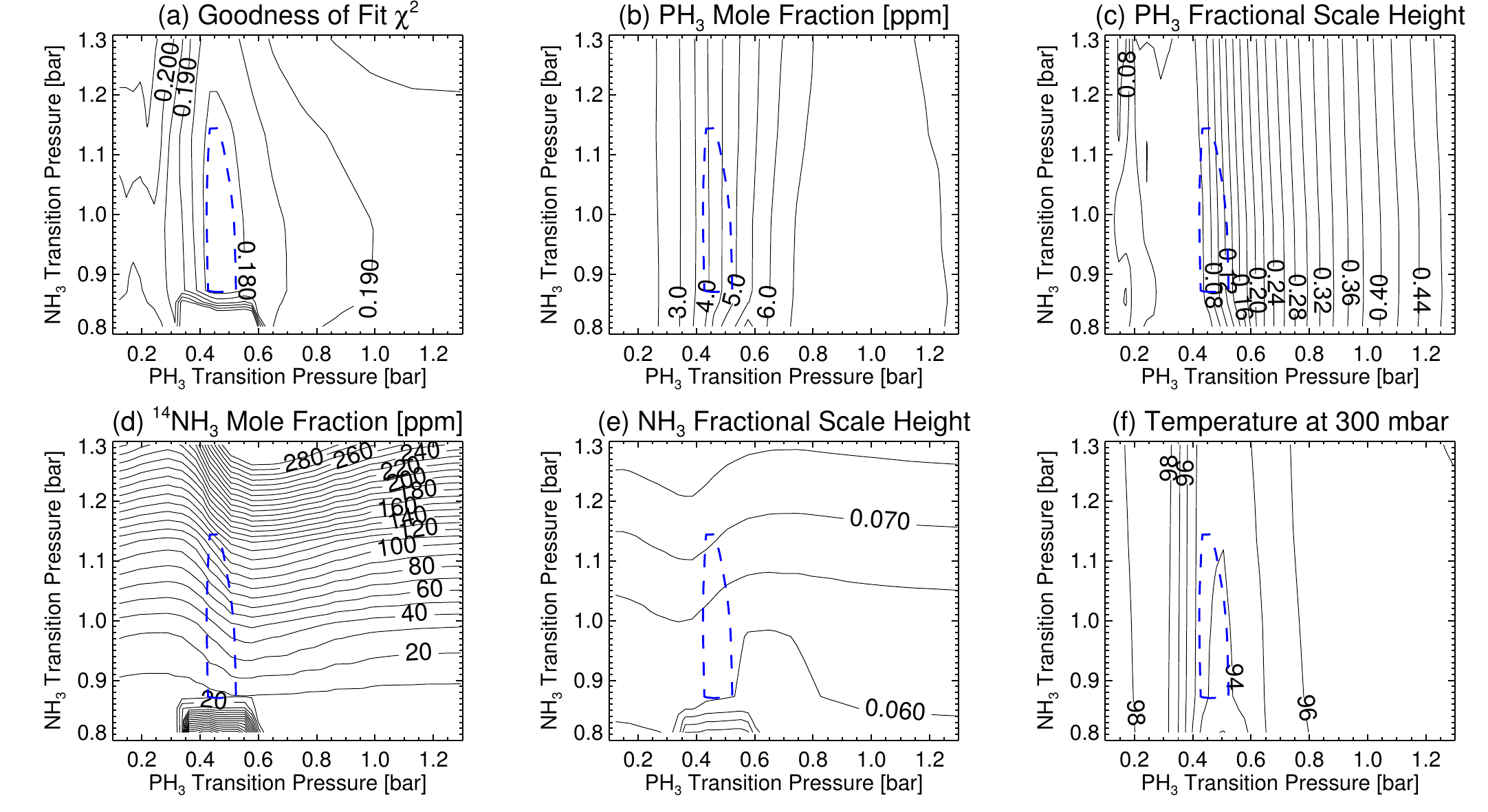}
\caption{Effect of varying the transition pressure $p_0$ for Saturn's ammonia and phosphine on the retrieved parameters.  Panel (a) indicates the small variations in fitting quality for the different knee assumptions and the $1\sigma$ uncertainty range for the two transition pressures (thick blue dashed contour).  Panels (b) and (c) show the variation in the PH$_3$ deep mole fraction and fractional scale height, respectively.  Panels (d) and (e) report the variations in deep mole fraction and fractional scale height for $^{14}$NH$_3$.  Panel (f) shows the retrieved 300-mbar temperatures.  This figure shows the case where the 900-\cm channel is scaled downward by a factor of 0.8 while the 960-\cm channel is unaltered.}
\label{sat_knee}
\end{center}
\end{figure*}

\begin{figure}[htbp]
\begin{center}
\includegraphics[width=8cm]{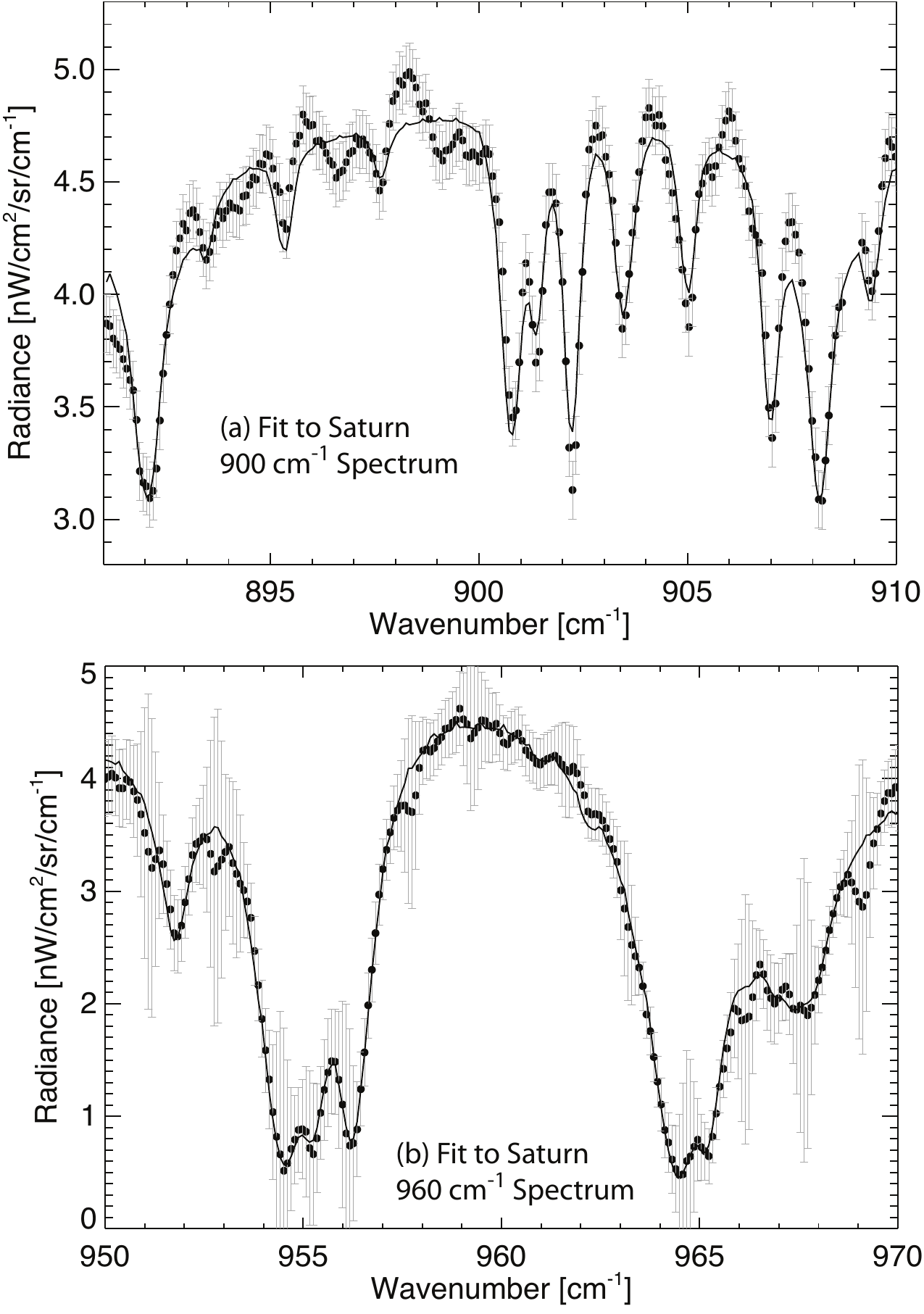}
\caption{Comparing the TEXES measurements (circles with error bars) and the best spectral fits (solid line) to the two Saturn spectra on February 3rd 2013.  The 900-\cm channel has been scaled by a factor of 0.8.  Error bars are inflated in regions of telluric contamination to ensure that the retrievals are weighted towards clear-sky regions of the spectrum.  Fits to the 900-\cm channel were improved by considering narrower sub-ranges, specifically omitting the 895-900-\cm region from the spectral fits which could be due to either poor telluric subtraction from the data or the absence of an unidentified emitter in our spectral model.}
\label{sat_bestfit}
\end{center}
\end{figure}

Fixing the 960-\cm channel as we did for Jupiter, we scaled the radiance in the 900-\cm channel and tested the spectral sensitivity to the PH$_3$ and NH$_3$ transition pressures, varying these between 300-1300 mbar for PH$_3$ and 800-1300 mbar for NH$_3$ in Fig. \ref{sat_knee}.  In general, the differences in the goodness-of-fit $\chi^2$ are small (Fig. \ref{sat_knee}a), and the sensitivity to the ammonia transition pressure is rather weak, with values in the 850-1150 mbar range satisfying the data to within $1\sigma$.  Lower transition pressures (850 mbar) required NH$_3$ mole fractions of 7 ppm and fractional scale heights of 0.06, whereas higher transition pressures (1100 mbar) required mole fractions of 135 ppm and fractional scale heights of 0.07, indicating the size of the acceptable solution space for NH$_3$ (Fig. \ref{sat_knee}d,e).   The PH$_3$ transition was rather better constrained to 400-550 mbar, which is at the lower limit of the pressure range provided by previous data analysis \citep{00orton, 07fletcher_ph3}.  The higher the radiometric scale factor for the 900-\cm channel, the lower the pressure of the PH$_3$ transition because there is less need for the broad PH$_3$ wings when the radiance is higher.  Intriguingly, only scale factors of 0.7-0.8 provide PH$_3$ transition pressures around 500-mbar that are consistent with previous studies, lending further support to our radiometric scaling.  Our best fit provided a PH$_3$ mole fraction of $4.1_{-0.2}^{+0.8}$ ppm with a fractional scale height of $0.08_{-0.02}^{+0.06}$, slightly smaller than those reported by Cassini/CIRS \citep{09fletcher_ph3} because the transition pressure is at a lower pressure (Fig. \ref{sat_knee}b,c).  Two examples of the quality of the spectral fits are shown in Fig. \ref{sat_bestfit}.   In general, the fit to the 960-\cm setting was superior to the fit to the 900-\cm setting, despite the additional terrestrial contamination in the 960-\cm channel.  Indeed, we were unable to fit precisely the peak-to-trough contrasts in the PH$_3$ lines in the 900-905-\cm region unless we ignored other regions of the spectrum, although the solution in Fig. \ref{sat_bestfit} is consistent with the observations within the random uncertainties.  The best-fitting $T(p)$ profile, PH$_3$ and $^{14}$NH$_3$ distributions for each scaling (five in total) will be used in Section \ref{sat_15N} to derive upper limits on the $^{15}$N/$^{14}$N ratio.




\subsection{Jupiter's $^{15}$N/$^{14}$N Ratio}
\label{jup_15N}

\begin{figure*}[htbp]
\begin{center}
\includegraphics[width=16cm]{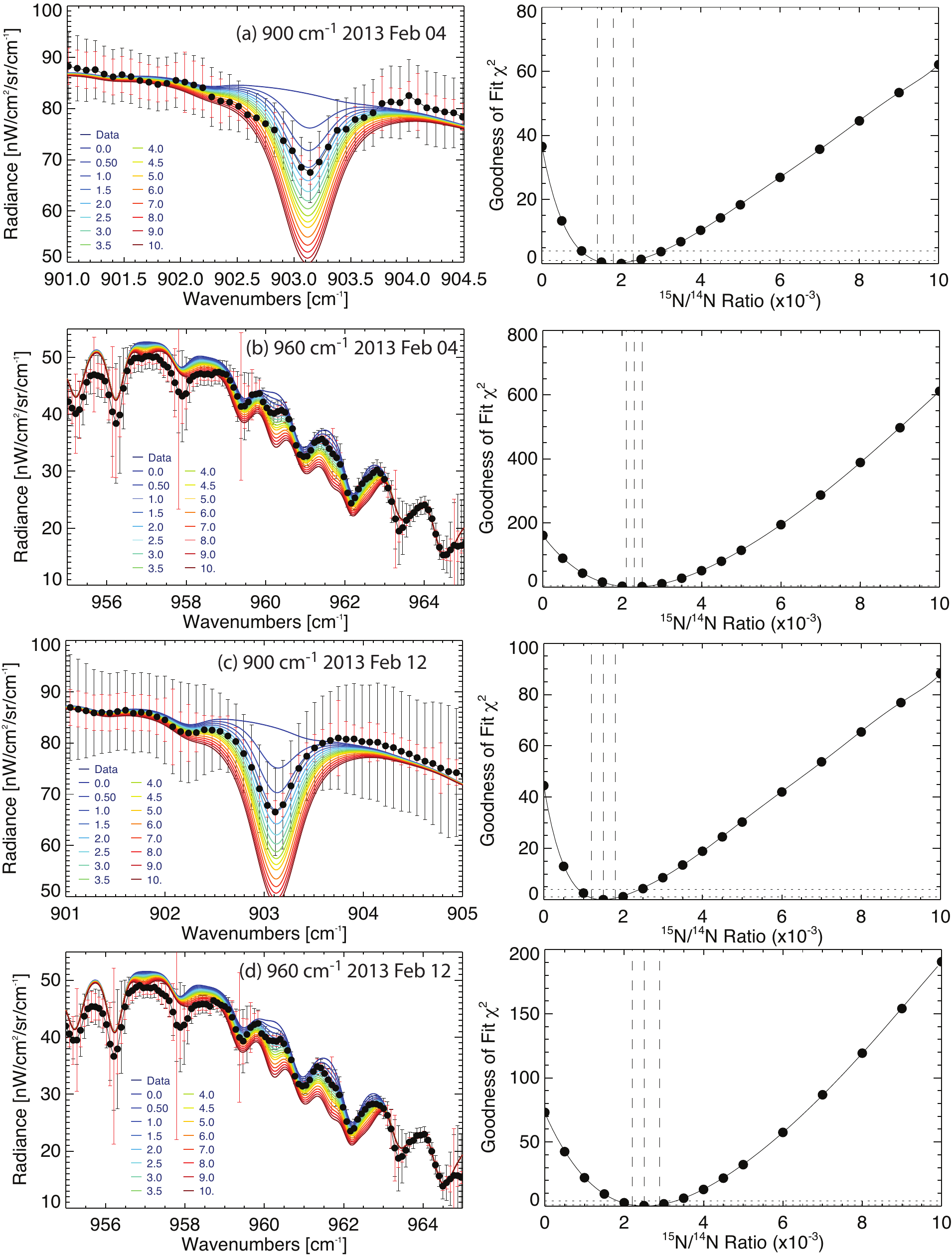}
\caption{Examples of the spectral fits to the Jupiter 900- and 960-\cm channels on February 4th and 12th 2013, using varying $^{15}$N/$^{14}$N ratios, and the $\chi^2$ goodness-of-fit for the best-fitting ratio (right hand column). In the left hand column, the data are the circular points, the coloured lines indicate models with different $^{15}$N/$^{14}$N ratios (in units of $10^{-3}$ according to the inset key).  Black error bars represent standard errors, red error bars indicate errors calculated with the modelled atmospheric transmission (see main text).  In the $\chi^2$ plots, the $1\sigma$ and $2\sigma$ limits are shown by horizontal dotted lines.  The optimal value and $1\sigma$ uncertainty is shown by vertical dashed lines.  Telluric contamination is indicated by the largest error bars. }
\label{jup15nh3}
\end{center}
\end{figure*}

Section \ref{jup_glob} modelled both the 900- and 960-\cm Jupiter observations simultaneously.  However, these observations were designed such that the $^{15}$N/$^{14}$N ratio could be derived from either one of these channels separately.  From this point on we use the `global solution' as the \textit{a priori} for a fit to a narrow sub-range surrounding the $^{15}$NH$_3$ features to provide the smallest possible residual while remaining consistent with the combined retrieval.  The $^{15}$NH$_3$ features were now omitted from the fitting between 902.6-903.8 \cm and 959-962 \cm, but we ensured that the $^{14}$NH$_3$ features were fitted as closely as possible.  The absolute calibration was varied as before to produce a new, refined grid of temperature, parameterised ammonia, aerosols and phosphine retrievals for each setting.  These retrieved properties were used to create forward models, scaling the $^{14}$NH$_3$ profile to provide $^{15}$N/$^{14}$N ratios between zero and $10\times10^{-3}$ (encompassing the Jupiter-like $2.3\times10^{-3}$ and Titan-like $6.0\times10^{-3}$ values).  Examples of the model-data comparisons and the goodness-of-fit $\chi^2$ are shown in Fig. \ref{jup15nh3}.  

In Section \ref{jup_glob} we used a comparison to CIRS-derived $T(p)$ from the 2000 flyby to restrict the region of parameter space under consideration.  The resulting 20-30\% decrease in the radiance measured in the 900-\cm channel was found to be consistent with the direct comparison of the CIRS and TEXES radiances.  Results for the $^{15}$N/$^{14}$N ratio in Table \ref{jupres} are therefore only quoted for the case where the radiometric scaling of the 960-\cm channel is left untouched, but the 900-\cm channel is reduced by factors of 0.7-0.8.  As the radiometric scaling strongly influences the retrieved temperatures, ammonia and aerosol opacity from each channel, the relatively narrow range of $^{15}$N/$^{14}$N solutions in Table \ref{jupres} is encouraging.

As described in Section \ref{model}, the $1\sigma$ uncertainty ranges were estimated via two techniques.  Firstly we used a 3\% random uncertainty in clear-sky spectral regions to replicate the standard error of coadded spectra (and arbitrarily large in regions of telluric contamination).  Secondly, we follow \citet{05greathouse} by evaluating $\chi^2$ using a scaling factor for the measured radiance based on the calculated terrestrial transmission ($\tau$), namely $\sigma=\sqrt(1.0-\tau)/\tau$.  Both approaches provide similar solutions but are included in Table \ref{jupres} for completeness.  Despite the $\pm20$\% radiometric uncertainty, Table \ref{jupres} shows that TEXES observations can confirm Jupiter's $^{15}$N/$^{14}$N ratio, a measurement only previously reported from space-based observatories.  Best-fitting values range from $1.4\times10^{-3}$ to $2.5\times10^{-3}$.  The 70\% scaling of the 900-\cm channel typically provides higher values of $^{15}$N/$^{14}$N than the 80\% scaling.  

A close inspection of the spectral fits suggests that the February 10, 2013 dataset was the least reliable.  Indeed, we struggled to fit the overall spectral shape with the model, so these ratios have correspondingly larger uncertainties.  A typical feature of the 960-\cm fit for all three dates (e.g., Fig. \ref{jup15nh3}b and c) was that the model failed to reproduce the 955-959 \cm region accurately, whereas 959-965 \cm was reproduced near-perfectly (the same discrepancy has been noted in modelling ISO data, T. Fouchet, \textit{pers. comms.}).  An additional absorber (e.g., aerosols) may be decreasing the measured radiance here, although no suitable candidates could be found.  Several telluric features are found in this area and could be reducing the quality of the fit.  Finally, Cassini/CIRS spectra were unable to shed light on the problem due to an electrical interference anomaly close to this problematic region \citep[e.g.,][]{06achterberg}.  Similarly, the slope of the continuum surrounding the 903-\cm $^{15}$NH$_3$ line was hard to reproduce in Fig. \ref{jup15nh3}a and c, with a tendency to overfit shorter wavenumbers and underfit longer wavenumbers. These offsets were folded into our error budget in Table \ref{jupres}.  In summary, TEXES observations are consistent with a jovian $^{15}$N/$^{14}$N ratio in the range from $1.4\times10^{-3}$ to $2.5\times10^{-3}$, provided that the Cassini/CIRS temperatures (2000) remain a valid assumption for the jovian tropospheric temperatures in 2013.  Independent constraints on either the temperature or the radiometric scale factor would improve this result in future studies.

\begin{sidewaystable}[htp]  
\caption{Range of solutions for Jupiter's $^{15}$N/$^{14}$N ratio, in units of $10^{-3}$, presented for each channel (either standard error or transmission error) for radiometric scalings of 0.7 and 0.8 for 900-\cm, and for 1.0 for the 960-\cm channel.  All values are quoted as $1\sigma$.}
\begin{center}
\begin{tabular}{|c|c|c|c|c|c|c|c|}
\hline
Channel & Radiometric & Feb 4 & Feb 4 & Feb 10 & Feb 10 & Feb 12 & Feb 12 \\
 & Scale & Standard & Transmission & Standard & Transmission & Standard & Transmission \\
  & & Error & Error & Error & Error & Error & Error  \\
\hline
900-\cm & 0.7 & $2.4^{+0.8}_{-1.0}$ & $2.3^{+0.7}_{-0.6}$ & $1.8^{1.6}_{-0.8}$ & $1.8^{+1.6}_{-0.8}$ & $2.0^{+1.1}_{-0.8}$ & $1.9^{+0.6}_{-0.4}$ \\

900-\cm & 0.8 & $1.9^{+0.7}_{-0.6}$ & $1.8^{+0.5}_{-0.4}$ & $1.4^{+1.5}_{-0.8}$ & $1.4^{+1.5}_{-0.8}$ & $1.5^{+0.8}_{-0.6}$ & $1.4^{+0.4}_{-0.3}$\\

960-\cm & 1.0 & $2.4^{+0.4}_{-0.4}$ & $2.3^{+0.2}_{-0.2}$ & $1.7^{+0.4}_{-0.3}$ & $1.7^{+0.5}_{-0.4}$ & $2.5^{+0.4}_{-0.3}$ & $2.5^{+0.2}_{-0.2}$ \\
\hline

\end{tabular}
\end{center}
\label{jupres}
\end{sidewaystable}%




\subsection{Saturn's $^{15}$N/$^{14}$N Ratio}
\label{sat_15N}
Having demonstrated that TEXES $^{15}$N/$^{14}$N ratio measurements on Jupiter were fully consistent with previous studies, we now proceed to estimate upper limits for the saturnian ratio.  Taking the best fitting global solution from Section \ref{satglob}, we refined the spectral fits in the vicinity of the $^{15}$NH$_3$ features for the two channels independently, repeating the retrieval within the $\pm20$\% systematic uncertainty envelope.  This step ensured a well-fit continuum prior to forward modelling of the $^{15}$NH$_3$ features.  The $^{15}$NH$_3$ profile was incorporated as a scaled version of the $^{14}$NH$_3$ profile determined in the previous stage. 

\begin{figure*}[htbp]
\begin{center}
\includegraphics[width=16cm]{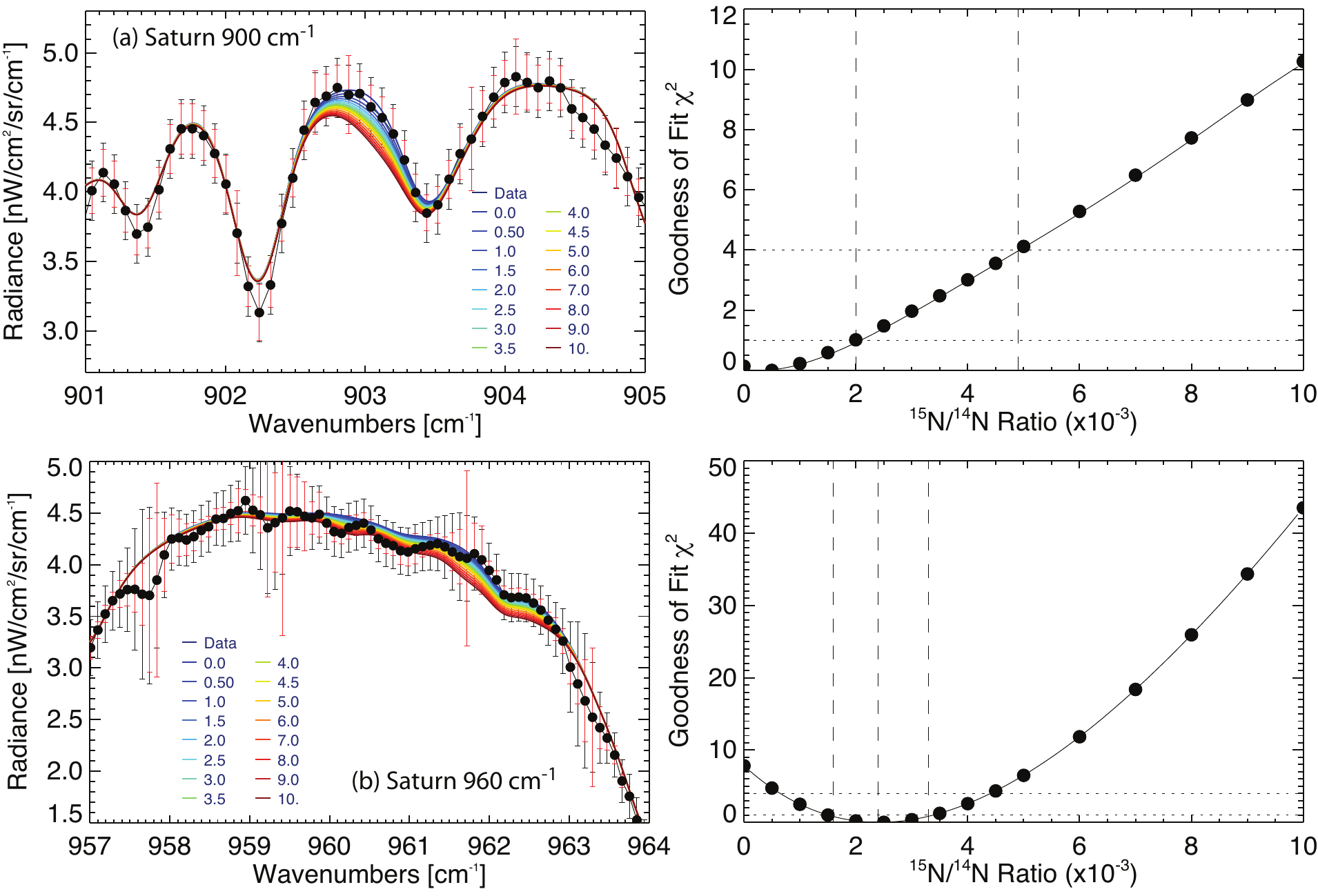}
\caption{An example of the quality of the spectral retrieval fit to the Saturn 900-\cm and 960-\cm channels on February 3rd 2013.  The 900-\cm channel has been scaled by a factor of 0.8. The left column shows the model fits (solid lines for different $^{15}$N/$^{14}$N ratios) to the data (points with error bars). Black error bars represent standard errors, red error bars indicate errors calculated with the modelled atmospheric transmission (see main text).  In the $\chi^2$ plots, the $1\sigma$ and $2\sigma$ upper limits are shown by horizontal dotted lines.  The apparent 'detection' in the 960-\cm channel is not considered robust given the quality of the spectral fit.  Telluric contamination is indicated by the largest error bars. }
\label{sat15nh3}
\end{center}
\end{figure*}

Table \ref{satres} reports the full solution space for each channel, uncertainty approach and radiometric scaling.  Trends are hard to identify, given that each radiometric scaling produced different \textit{absolute} values for the tropospheric temperature and ammonia abundances.  The $^{15}$N/$^{14}$N ratio is generally permitted to be larger in the 960-\cm channel because of the greater effect of telluric contamination (see the spectral fit quality in Fig. \ref{sat15nh3}).  If we assume that the radiometric calibration of the 960-\cm channel is accurate, and that the 900-\cm channel was more easily fit if reduced by 20\% (which provided a PH$_3$ distribution most similar to previous investigations in Section \ref{satglob}), then we conclude from Fig. \ref{sat15nh3} that $^{15}$N/$^{14}$N$<2.0\times10^{-3}$ for the 900-\cm channel and $^{15}$N/$^{14}$N$<2.8\times10^{-3}$ for the 960-\cm channel (assuming standard errors).  Only the largest value in our solution space ($<3.6\times10^{-3}$ for a scaling of the 960-\cm channel by 0.8) is consistent with the terrestrial value, with all other solutions being smaller and more consistent with the jovian value.   It should be noted that these are $1\sigma$ upper limits, representing a 68.3\% probability that the $^{15}$N/$^{14}$N ratio falls within this range.  A more conservative upper limit of $2\sigma$ (95.4\% of the probability) limits the ratio to $^{15}$N/$^{14}$N$<4.9\times10^{-3}$ for the 900-\cm channel and $^{15}$N/$^{14}$N$<4.5\times10^{-3}$ for the 960-\cm channel (Fig. \ref{sat15nh3}), which remains smaller than the Titan value of $6\times10^{-3}$.   A $3\sigma$ upper limit permits all solutions in our tested range.

\begin{sidewaystable}[htp]  
\caption{Range of solutions for Saturn's $^{15}$N/$^{14}$N ratio, in units of $10^{-3}$, presented for each channel and uncertainty estimate (either standard error or transmission error) for a range of radiometric scalings.  All upper limits are quoted as $1\sigma$.}
\begin{center}
\begin{tabular}{|c|c|c|c|c|}
\hline
Radiometric & 900-\cm & 900-\cm & 960-\cm  & 960-\cm  \\
Scaling & [Standard Error] &[Transmission Error] & [Standard Error] & [Transmission Error] \\
\hline
0.8 & $<2.0$ & $<1.7$ & $<3.6$ & $<3.4$ \\
0.9 & $<2.5$ & $<2.1$ & $<2.6$ & $<2.5$ \\
1.0 & $<1.5$ & $<1.1$ & $<2.8$ & $<2.7$ \\
1.1 & $<1.7$ & $<1.2$ & $<1.9$ & $<1.8$ \\
1.2 & $<3.3$ & $<2.7$ & $<2.8$ & $<2.8$ \\
\hline
\end{tabular}
\end{center}
\label{satres}
\end{sidewaystable}%

\subsubsection{Medium resolution Saturn observations}
The low-resolution nodded spectra in Fig. \ref{sat15nh3} were both obtained on the same date (February 3, 2013), but we sought to confirm these upper limits by comparing to spectra obtained with a medium-resolution TEXES mode ($R\approx10000$), utilising 900-\cm spectra from February 2, 2013 and 960-\cm spectra from January 16, 2012 (see Section \ref{data}).  These were not optimised for the $^{15}$NH$_3$ study as they failed to perform a simultaneous measurement of $^{14}$NH$_3$ features to extract a reliable ratio.  Nevertheless, we could fix the $^{14}$NH$_3$ abundances to those derived from the low-resolution spectra, and fit the medium-resolution spectra to check consistency with our derived upper limits.  

\begin{figure}[htbp]
\begin{center}
\includegraphics[width=8cm]{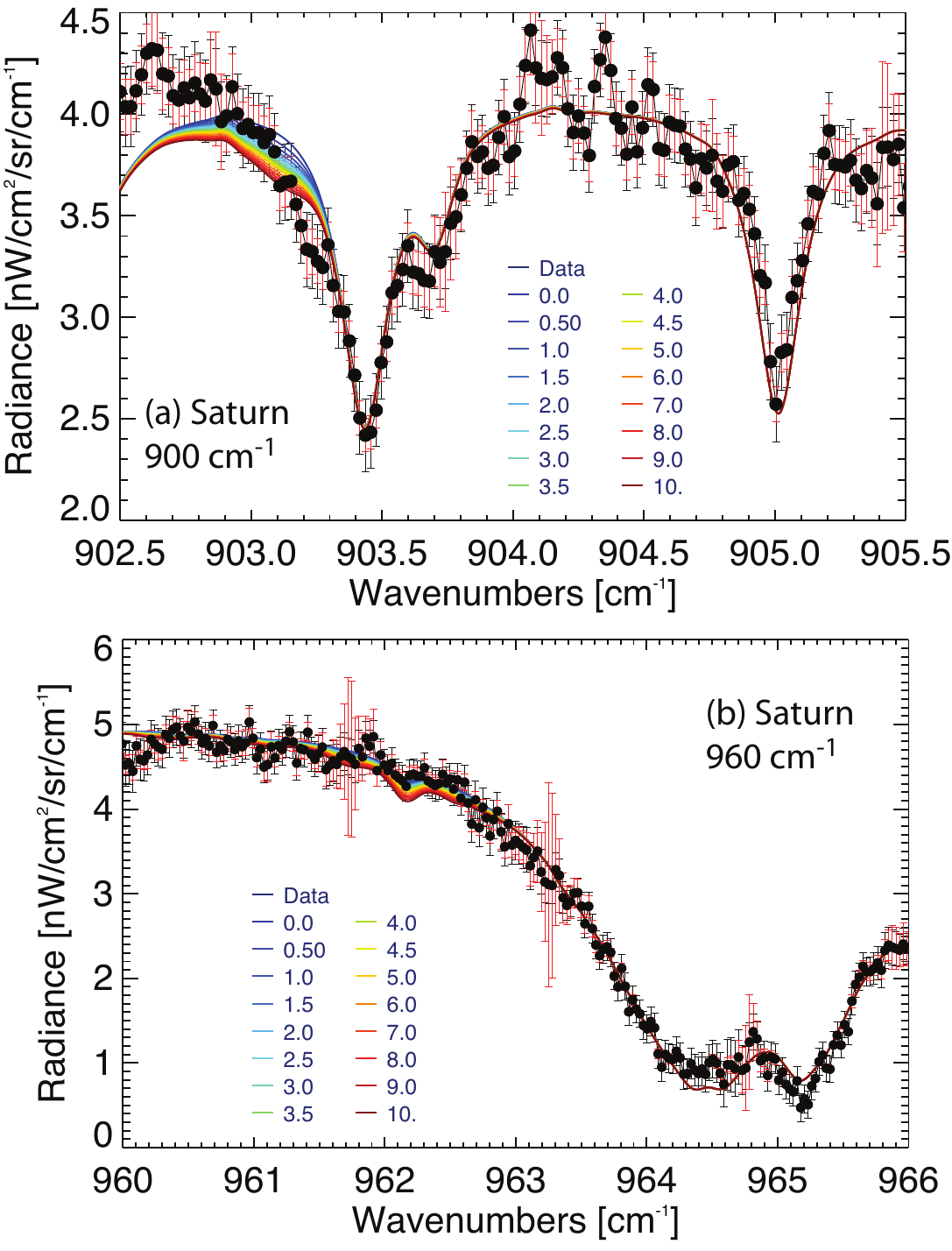}
\caption{Spectral fits to the medium-resolution TEXES spectra of Saturn on January 16th 2012 (960-\cm channel, panel b) and February 2nd 2013 (900-\cm channel, panel a).  Continuum noise in the 900-\cm channel prevented extraction of reliable upper limits.  Upper limits derived from the 960-\cm channel were consistent with those derived from the low-resolution nodded observations, despite the weakness of this spectral feature.  }
\label{medres_fit}
\end{center}
\end{figure}

$k$-distribution tables were produced for the medium resolution settings and used to fine-tune the TEXES wavelength calibration.  Spectral fits were generated by retrieving tropospheric temperatures and a parameterised PH$_3$ distribution for radiometric scalings of $\pm20$\% around the TEXES calibration, and upper limits derived accordingly.  The quality of the data and the best-fit spectral model are shown in Fig. \ref{medres_fit}.  For the 900-\cm channel the deep PH$_3$ mole fraction varied from 7.0-7.2 ppm depending on the radiometric scaling, whereas the 960-\cm channel had best fits from 5.6-6.4 ppm one year earlier.  Although temporal evolution of PH$_3$ is plausible (especially given the dynamics of Saturn's 2010-11 storm), systematic offsets in calibration are more likely.  From the 900-\cm spectrum in Fig. \ref{medres_fit}a it quickly became obvious that the continuum noise would hamper any attempts at deriving a reasonable $^{15}$NH$_3$ upper limit, despite the good fits to the PH$_3$ lines themselves.  Indeed, a $\chi^2$ test suggested that all $^{15}$N/$^{14}$N ratios smaller than $7.5\times10^{-3}$ were statistically permissible.  Longer integrations would be necessary to reduce the continuum noise in the medium-resolution 900-\cm channel.  

Fits to the 960-\cm medium-resolution channel were more promising as the continuum in Fig. \ref{medres_fit}b was much smoother.  Although no specific $^{15}$NH$_3$ features were detected, the data favour Jupiter-like $^{15}$N/$^{14}$N ratios rather than Titan-like ones.  Assuming that the TEXES calibration is accurate, we find $1\sigma$ optimal values of $(2.5\pm1.5)\times10^{-3}$ for the Saturn ratio (note that we do not claim this as a detection given the ambiguity of features in Fig. \ref{medres_fit}b).  Alternatively, varying the radiometric scale by $\pm20$\% yields upper limits of $<3.2\times10^{-3}$ and $<2.3\times10^{-3}$ for radiometric scalings of 80\% and 120\%, respectively (note that this assumes standard errors).  However, these estimates are rather crude because $^{14}$NH$_3$ could not be simultaneously constrained, and relies on the low-resolution estimate from 2013.  In summary, the medium-resolution TEXES observations of Saturn provide a useful consistency check and support a Jupiter-like low $^{15}$N enhancement, but are insufficient alone to constrain the $^{15}$N/$^{14}$N ratio. 

\subsection{Latitudinally-resolved retrievals}
\label{zonal}

\begin{figure*}[htbp]
\begin{center}
\includegraphics[width=16cm]{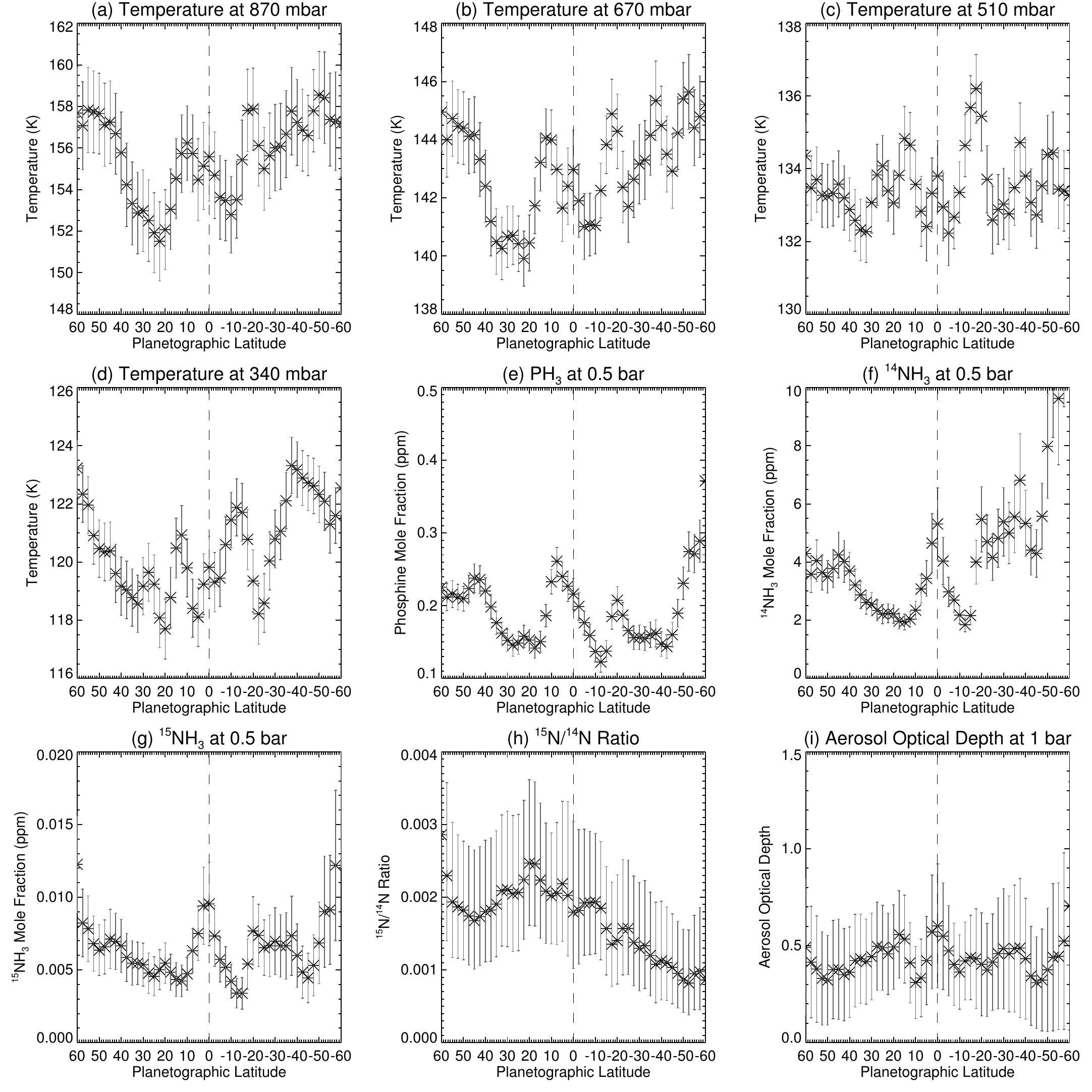}
\caption{Zonally resolved retrievals of Jupiter's tropospheric temperature and gaseous composition from low-resolution TEXES spectra on February 4th 2013, using a combination of 900-\cm and 960-\cm channels.  Panels (a)-(d) depict temperatures at four pressure levels in the atmosphere; panels (e)-(g) show the 500-mbar abundances of PH$_3$, $^{14}$NH$_3$ and $^{15}$NH$_3$; panel (h) shows the $^{15}$N/$^{14}$N ratio and (i) shows the cumulative 10-$\mu$m aerosol opacity at 1 bar.  Absolute quantities are sensitive to the systematic uncertainties in assumed radiometric calibration, whereas spatial contrasts are expected to be robust.}
\label{jupmerid}
\end{center}
\end{figure*}

\begin{figure*}[htbp]
\begin{center}
\includegraphics[width=16cm]{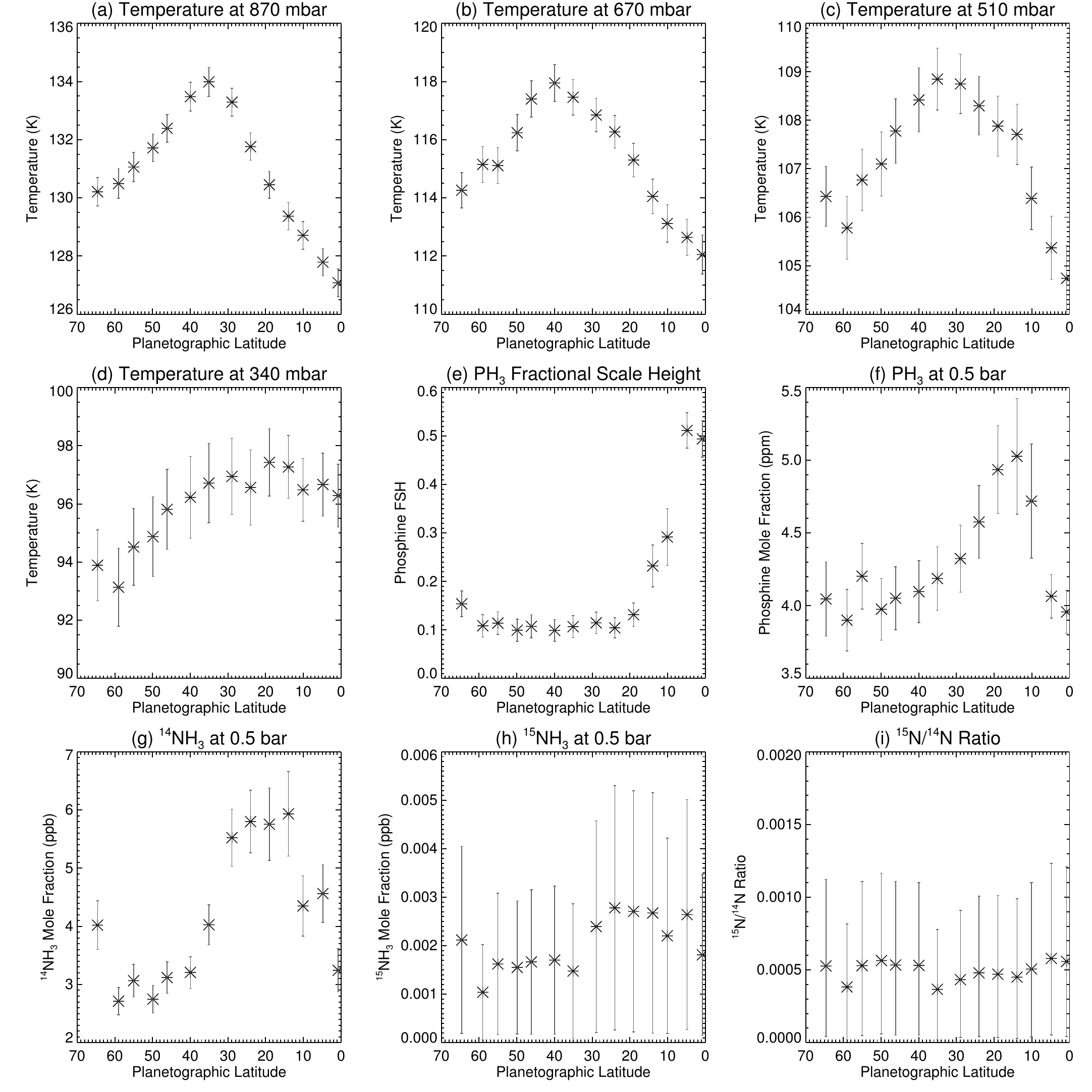}
\caption{Zonally resolved retrievals of Saturn's tropospheric temperature and gaseous composition from low-resolution TEXES spectra on February 3rd 2013, using a combination of 900-\cm and 960-\cm channels.  Panels (a)-(d) depict temperatures at four pressure levels in the atmosphere; panels (e)-(f) show two quantities parameterising the PH$_3$ distribution; panels (g)-(h) provide scale factors for \textit{a priori} profiles of two ammonia isotopologues; and panel (i) shows the $^{15}$N/$^{14}$N ratio.  Absolute quantities are sensitive to the systematic uncertainties in assumed radiometric calibration, whereas spatial contrasts are expected to be robust.}
\label{sat_merid}
\end{center}
\end{figure*}

The final step in our analysis sought to extract zonal mean temperatures and abundances from the Jupiter scans and the Saturn nodded spectra.  Spectra were binned on a Nyquist-sampled $5^\circ$ latitude grid.  The 900-\cm and 960-\cm channels were fitted simultaneously, scaling the 900-\cm channel by a factor of 0.8.  The fitting provided retrieved mole fractions of $^{15}$NH$_3$ rather than manually adjusting the $^{15}$N/$^{14}$N ratio for every spectrum; this made the process more computationally efficient, but had the effect of sacrificing accuracy. In general, the zonally-resolved spectra were more noisy than the global average analysed above, such that constraints on the gaseous abundances and $^{15}$N/$^{14}$N ratio are worsened.  Nevertheless, this experiment demonstrates the capability for TEXES to derive spatially-resolved atmospheric properties.  Absolute values in Figs. \ref{jupmerid} and \ref{sat_merid} are sensitive to the assumed radiometric calibration uncertainty of $\pm20$\%, so only the relative variations in these parameters can be taken as robust.

For Jupiter we simultaneously retrieved tropospheric temperatures and scaled profiles of aerosols (a compact cloud with an 800-mbar cloud base), phosphine (based on the vertical profile of J. Moses, \textit{pers. comms.} and the two ammonia isotopologues.  The \textit{a priori} $^{14}$NH$_3$ profile was based on a mean of the global results in Table \ref{jupcomp}, and scaled by a factor of $2\times10^{-3}$ to provide the \textit{a priori} $^{15}$NH$_3$ profile.  Fig. \ref{jupmerid} shows latitudinal variations of temperature, aerosol optical depth at 10 $\mu$m, PH$_3$ and the two isotopologues of NH$_3$ on February 4th 2013 (similar meridional trends were produced on all three dates).  Tropospheric temperatures (Fig. \ref{jupmerid}a-d) show the contrasts of the SEB, equator and NEB as inferred from the raw images in Fig. \ref{images}, and differ moderately from CIRS derivations \citep{06simon,06achterberg} in that the typically cold equatorial zone features a moderately-warm central band, possibly associated with a narrow red haze visible during this period.  Aerosol opacity (Fig. \ref{jupmerid}i) is largely homogeneous with latitude, although we caution that the 900- and 906-\cm channels have only limited sensitivity to this parameter.  The latitudinal distribution of PH$_3$ is qualitatively similar to \citet{09fletcher_ph3}, but the 500-mbar mole fraction is smaller by a factor of two, likely due to systematic offsets in the radiometric calibration.   The distribution of $^{14}$NH$_3$ (Fig. \ref{jupmerid}f) shows a sharper equatorial peak but similar latitudinal trends and mole fractions to \citet{06achterberg}, with the exception of latitudes poleward of $20^\circ$S, where higher mole fractions give the appearance of a hemispheric asymmetry in NH$_3$.  The source of this asymmetry is unclear, but it was reproduced in tests with a variety of different retrieval assumptions.  The asymmetry is less pronounced in the $^{15}$NH$_3$ distribution (Fig. \ref{jupmerid}g), and hence the $^{15}$N/$^{14}$N ratio (Fig. \ref{jupmerid}h) shows a gradient from north to south (this gradient was observed on all three dates). Quantitatively, the global value of the $^{15}$N/$^{14}$N ratio is the same as that from the averaged spectrum analysed in Section \ref{jup_15N}.  The meridional ammonia and phosphine distributions will be the topics of future studies incorporating temperature and aerosol constraints from additional TEXES channels.

For Saturn we retrieved tropospheric temperatures, a parameterised PH$_3$ profile and simple scalings of the $^{14}$NH$_3$ and $^{15}$NH$_3$ profiles.  For our \textit{a priori} $^{14}$NH$_3$ profile we assumed 100 ppm for $p>950$ mbar and a fractional scale height of 0.08; scaling it by $1\times10^{-3}$ for the \textit{a priori} $^{15}$NH$_3$ profile.  The results are shown in Fig. \ref{sat_merid}, which shows the latitudinal trends for the temperature and composition.   Despite the uncertainties in absolute quantities, Fig. \ref{sat_merid} shows relative spatial variations that are real.  Zonal temperatures peak between 30-40$^\circ$N in the latitude band associated with the 2010-11 storm system \citep{11fletcher_storm}, and the 3-4 K contrast between the storm band and adjacent latitudes is consistent with the Cassini/CIRS findings of \citet{14achterberg}.  More importantly, neither PH$_3$ nor NH$_3$ show peaks in this band, confirming that the retrieval of temperature remains independent from the gaseous abundances.  The PH$_3$ fractional scale height (Fig. \ref{sat_merid}e) rises towards the equator to produce a meridional profile consistent with Cassini/CIRS \citep{09fletcher_ph3}, whereas the equatorial drop in the deep PH$_3$ mole fraction (Fig. \ref{sat_merid}f) was also observed by Cassini/VIMS \citep{11fletcher_vims}.  The $^{14}$NH$_3$ profile (Fig. \ref{sat_merid}g) deviates from the meridionally-uniform profile derived from Cassini/CIRS \citep{12hurley} and from the equatorial maximum detected by Cassini/VIMS \citep{11fletcher_vims}.  However, we cannot rule out temporal variations associated with Saturn's springtime storm and seasonally-evolving haze distributions, and the implications for this TEXES-derived ammonia distribution remain unclear.  These latitudinally-resolved spectra offer negligible constraint on the $^{15}$NH$_3$ distribution, but the retrieved $^{15}$N/$^{14}$N ratios are consistent with the low 'Jupiter-like' values and inconsistent with 'Titan-like' values, as described in Section \ref{sat_15N}.  Finally, given the random uncertainty on these spectra we see no evidence for variability in the ratio with latitude (i.e., no dynamical or seasonal influences are apparent in these results).

\section{Discussion:  The Origin of Giant-Planet Nitrogen}
\label{discuss}

\begin{figure*}[htbp]
\begin{center}
\includegraphics[width=10cm]{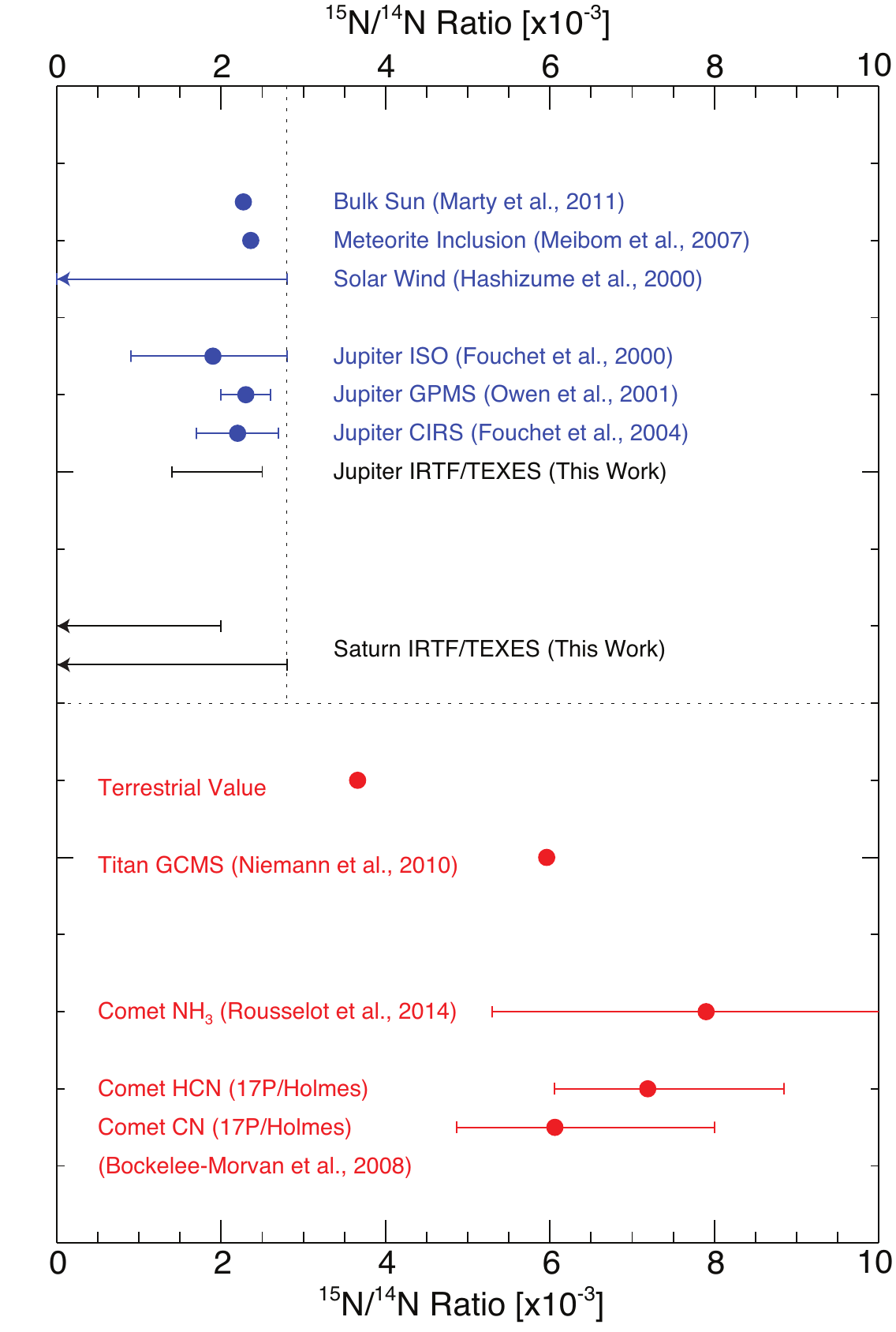}
\caption{Comparison of the $^{15}$N/$^{14}$N ratio on Jupiter and Saturn from this work with previous studies of Jupiter \citep{00fouchet, 01owen, 04fouchet}, the solar environment \citep{11marty, 07meibom, 00hashizume}; Earth and Titan \citep{10niemann}; and a representative sample of cometary $^{15}$N/$^{14}$N ratios in NH$_3$ \citep{14rousselot}, HCN and CN \citep{08bockelee-morvan}.  The horizontal dashed line separates 'primitive' solar nebula nitrogen (blue points) from nitrogen compounds found in comets, or nitrogen in secondary atmospheres (Earth, Titan, red points).}
\label{Nratio}
\end{center}
\end{figure*}

The low-spectral resolution TEXES observations of Jupiter and Saturn provided (i) ground-based confirmation of the low $^{15}$N-enrichment of Jupiter's nitrogen inventory; and (ii) the first upper limits on Saturn's $^{15}$N-enrichment, suggesting a $^{15}$N/$^{14}$N ratio more consistent with that of Jupiter than with Earth or Titan-like values.  Fig. \ref{Nratio} compares these giant planet values to those found elsewhere in our solar system.  The primary result of this study is the similarity of the $^{15}$N/$^{14}$N ratios for Jupiter and Saturn (assuming $1\sigma$ uncertainties), and the consistency with primordial solar nebula values inferred from the solar wind (see Section \ref{intro}).  What does this similarity imply for the cosmochemical source reservoirs for the nitrogen-carrying species (ices, gases) accreted into the proto-Jupiter and proto-Saturn?  We described in Section \ref{intro} that a low $^{15}$N enrichment favoured accretion of primordial N$_2$, rather than condensed nitrogen molecules (NH$_3$, HCN, CN, etc.) that would be enriched in $^{15}$N via isotopic exchanges in ion-molecule reactions \citep[e.g.,][]{00terzieva}, but does this help differentiate between models of giant planet origins?

To explain Jupiter's solar balance of elemental enrichments, in addition to the low $^{15}$N enrichment, \citet{95owen,01owen,03owen} argue for trapping of N$_2$ and other volatiles in porous amorphous ices at very low temperatures (i.e., at large heliocentric distances within the interstellar cloud).  However, volatile trapping in crystalline ices, or condensation of pure ices, are also able to reproduce the solar balance of Jupiter's enrichments \citep[e.g.,][and references therein]{09mousis}, meaning that we must rely on additional information to distinguish these theories.  The limited remote sensing information available for Saturn \citep[e.g.,][]{08hersant, 14mousis} suggests a non-solar balance of elemental enrichments.  Indeed, the C/N ratio on Jupiter was shown to be solar to within error bars, whereas it was tentatively supersolar on Saturn to within measurement uncertainties (i.e., Saturn's nitrogen enrichment is lower than its carbon enrichment), implying that the source reservoirs for both planets are not necessarily the same.  Here we assume that the maximum NH$_3$ abundances reported from millimetre and 5-$\mu$m spectroscopy of Saturn \citep[approximately 500 ppm,][]{85depater,11fletcher_vims} are representative of the deep sub-cloud reservoir of nitrogen, although this measurement must ultimately be confirmed via \textit{in situ} probing. If both planets formed from a reservoir of cold-trapped N$_2$ in amorphous ices, we might expect a solar balance of enrichments on Saturn too, which does not appear to be the case.

At the location of giant-planet formation, the disc temperatures are expected to have been too warm for significant adsorption of volatiles on amorphous ices \citep{01gautier}.  Instead, volatiles could have been trapped by crystalline water ice in the form of clathrates or hydrates during the slow cooling of the nebula \citep{85lunine, 05gautier, 01gautier,04hersant, 05alibert, 06mousis, 08hersant}.  Clathrates sequester guest gases within a hydrogen-bonded water ice structure.   In this scenario, the balance of the different elements depends on the efficiency of enclathration and the availability of crystalline water ice \citep{09mousis}, which is rather poorly understood in the early solar nebula.  A deficiency of crystalline water ice at 10 AU may have inhibited the formation of N$_2$ clathrate in Saturn's feeding zone \citep{04hersant, 05gautier}, and warm formation temperatures could have prevented direct N$_2$ condensation \citep{08hersant}, leading to a saturnian C/N ratio that is supersolar. \citet{05gautier} and \citet{08hersant} therefore predicted that Saturn should have a higher $^{15}$N-enrichment than that observed in Jupiter due to the increased importance of primordial NH$_3$ in Saturn's feeding zone, which our new upper limit no longer permits.  Conversely, \citet{09mousis} questioned the reliability of Saturn's measured nitrogen enrichment, and included pure N$_2$ condensate formation at cold temperatures (i.e., after the disc had cooled sufficiently over time) to predict the elemental complements of Jupiter and Saturn. By allowing N$_2$ to have a more important contribution to the proto-Saturn, the model of \citet{09mousis} would decrease the hypothesised $^{15}$N enrichment, but predictions of the bulk nitrogen enrichment remain to be tested.

%


Could trapped ammonia have played a role in Saturn's feeding zone?  Accretion of primordial NH$_3$, and subsequent production of a secondary atmosphere from the original carrier by photolysis \citep{78atreya}, has been invoked as a possible explanation for Titan's high $^{15}$N/$^{14}$N ratio of $(5.96\pm0.02)\times10^{-3}$ \citep{10niemann}, which is considerably larger than that found on Jupiter, Saturn (to $2-\sigma$ confidence) or the Earth (Fig. \ref{Nratio}).  Although Titan's enrichment could be related to fractionation by atmospheric escape processes \citep[e.g., see reviews of Titan's origins by ][]{10lunine,10atreya}, \citet{09mandt} suggest that over 200 bars of Titan's atmosphere would need to have been lost, and that the present-day $^{15}$N/$^{14}$N ratio reflects that acquired from ammonia-ice accretion.  Accretion of a $^{15}$N-enriched compound could explain both the $^{15}$N enrichment and the low $^{36}$Ar/N$_2$ ratio in Titan's atmosphere \citep[if Titan's nitrogen was from primordial N$_2$, the low condensation temperatures required would have been sufficient to provide argon and nitrogen in a solar-like ratio,][]{82owen}.  However, as reviewed by \citet{10lunine, 10atreya}, other mechanisms (such as adsorption onto photochemical hazes in Titan's atmosphere) might be responsible for the depletion of $^{36}$Ar.   Nevertheless, this Titan accretion scenario, along with the difficulty in trapping N$_2$, would seem to favour accretion of Titan from a $^{15}$N-enriched source reservoir (such as primordial NH$_3$). 

This TEXES upper limit on $^{15}$N enrichment, along with the observation that Saturn's nitrogen is indeed enriched over solar composition \citep[see ][for a recent review, and references therein]{14mousis}, argue against delivery from a primordial reservoir of NH$_3$ and in favour of N$_2$ as the dominant carrier.  Indeed, the N$_2$ reservoir dominated over the NH$_3$ reservoir for the feeding zones of both planets.  This leaves us with a conundrum for solid-planetesimal accretion - how might icy materials rich in N$_2$ survive being transported from the distant, low-temperature reaches of our solar system ($T_f<40$ K for N$_2$ trapping in both amorphous and crystalline scenarios), in through the disc to the location of giant planet accretion?  The disc would cool with time, so N$_2$-rich ices could be preserved if the timescales were long enough, but the dynamics of the planetesimal population and the timescale for giant planet formation remain poorly understood \citep[e.g., see the recent review by][]{13helled}.  The question of \textit{how} a forming giant planet could accrete solid ices (either amorphous or crystalline) from the distant, cold solar system without devolatilisation as they approach the feeding zones of the proto-Jupiter and proto-Saturn remains unresolved \citep{05gautier}.  

A final possibility is the accretion of N$_2$ as a gas directly from the solar nebula along with hydrogen and helium during the runaway accretion phase.  Given that N$_2$/H$_2$ is expected to be larger than NH$_3$/H$_2$ in the solar nebula \citep{80lewis}, this scenario would provide the low $^{15}$N enrichment to be consistent with the Jupiter and Saturn observations.  NH$_3$ or other nitrogen compounds trapped in crystalline ices or comets could not have been a major contributor to the nitrogen inventory of either planet (although NH$_3$ must have played some role at 10 AU to explain Titan's inventory).  However, a mechanism must be identified to enhance the nitrogen inventory over solar composition on both planets.  It should be noted that the same problem of elemental enrichment is faced by the disc instability mechanism of giant planet formation, but \citet{13helled} review possible mechanisms of producing super-solar enhancements in forming proto-planets (local enhancements in spiral arms of a disc, planetesimal capture after formation, and tidal stripping).  Their message is that supersolar elemental enrichments alone should not be taken as evidence of the core accretion model.  It might be possible that accretion of clathrated N$_2$ played a more important role at Jupiter where the availability of crystalline water ice was greater, explaining the solar C/N ratio on Jupiter compared to the supersolar C/N ratio on Saturn.  In reality, the bulk composition of Jupiter and Saturn likely originated from some combination of volatile trapping in ices, condensation of pure ices and gas phase accretion.  The relative importance of each contribution will be the subject of future modelling studies.

\section{Conclusions}
\label{conclude}

Low-resolution, mid-infrared spectral mapping of Jupiter and Saturn in 2013 using the TEXES instrument on the IRTF has permitted the first ground-based study of the $^{15}$N/$^{14}$N ratio on the two gas giants (Fig. \ref{Nratio}).  Observations focussed on two spectral channels ($\pm20$ \cm surrounding 900- and 960-\cm) selected for (i) nearby signatures of $^{14}$NH$_3$ and $^{15}$NH$_3$ to allow simultaneous measurement; (ii) close proximity to PH$_3$ features to characterise the spectral continuum; and (iii) limited contamination from telluric features.  Jupiter was observed in a scan mode to map the spatial variability of emission in each channel, whereas Saturn was observed in nod mode to determine the latitudinal variability.  Given the higher radiance, Jupiter observations were repeated on three nights to compare the precision of calibration (with the caveat that precision does not imply radiometric accuracy).   

The primary conclusion of this research is that Jupiter and Saturn's $^{15}$N/$^{14}$N ratios are indistinguishable from one another, and that both are consistent with the primordial $^{15}$N/$^{14}$N ratio in the solar nebula.  $^{15}$NH$_3$ features were readily detectable on Jupiter, yielding isotope ratios ranging from $1.4\times10^{-3}$ to $2.5\times10^{-3}$ that are consistent with both the \textit{in situ} result from the Galileo Probe Mass Spectrometer \citep[$(2.3\pm0.3)\times10^{-3}$,][]{01owen} and the remote sensing result from Cassini \citep[CIRS, $(2.2\pm0.5)\times10^{-3}$,][]{04fouchet}.  This represents the first ground-based determination of this ratio.  Conversely, we were unable to detect $^{15}$NH$_3$ features in Saturn's phosphine-dominated spectrum.  Nevertheless, we report the first $1\sigma$ upper limits on Saturn's $^{15}$N/$^{14}$N ratio, requiring ratios no larger than $2.0\times10^{-3}$ for the 900-\cm channel and $2.8\times10^{-3}$ for the 960-\cm channel (the second channel providing a consistent, but less stringent, upper limit).  The values were supported by medium-resolution spectroscopy acquired a year earlier in 2012.  A more conservative $2\sigma$ confidence limit raises this upper limit to a $^{15}$N/$^{14}$N ratio less than $5\times10^{-3}$.  Specifically, we find that a Titan-like $^{15}$N/$^{14}$N ratio \citep[$6.0\times10^{-3}$,][]{10niemann} is not supported by the Saturn observations, implying that $^{15}$N-enriched primordial ammonia could not have provided a substantial contribution to Saturn's nitrogen inventory.

The dominant uncertainty in this ground-based study is the radiometric calibration.  Although TEXES absolute calibrations are consistent from night to night (Section \ref{data}), comparisons of the Jupiter observations with Cassini/CIRS data in 2000, combined with cross-checks of the tropospheric temperatures and vertical PH$_3$ distributions with previous studies, all suggest that the 900-\cm flux should be consistently decreased by 20-30\% relative to the quoted flux calibration.  A $\pm20$\% radiometric uncertainty admits a large solution space for temperatures and gaseous composition, and it was only by imposing additional constraints (i.e., temperatures and phosphine could not deviate substantially from previous studies) that we could determine the atmospheric composition.   This was somewhat alleviated in this study by simultaneously observing lines of both ammonia isotopologues and then comparing ratios for a range of different radiometric scalings.  Future studies must either (i) reduce the radiometric uncertainty or (ii) provide independent temperature constraints in order to improve the accuracy of these ground-based measurements.

Much of this study focussed on obtaining close fits to spatial averages of the Jupiter and Saturn data to derive the $^{15}$N/$^{14}$N ratio, but zonal mean atmospheric properties (tropospheric temperatures, aerosol opacity and distributions of PH$_3$ and NH$_3$) were presented in Section \ref{zonal} as a secondary product of this analysis, highlighting the capabilities of TEXES for spatial mapping of gas giant tropospheric conditions.  TEXES scan mapping of Jupiter provided zonal mean temperatures consistent with previous studies (i.e., belt/zone contrasts bounded by Jupiter's zonal jets) and revealed dramatic wave structure in Jupiter's NEB in 2013.  The cold equatorial zone featured a warm, diffuse band in February 2013, coincident with a faint red haze observed by amateur astronomers during that period.  Oval BA was characterised by both a cold central vortex and a warm peripheral ring related to secondary circulation in the anticyclone.   Ammonia and phosphine were both elevated in Jupiter's equatorial zone above the clouds, with abundances similar to those derived from Cassini/CIRS \citep{06achterberg, 09fletcher_ph3}.  Aerosol optical depth was poorly constrained by these TEXES channels.  Jupiter's $^{15}$N/$^{14}$N ratio consistently showed a hemispheric asymmetry on all three nights (related to elevated $^{14}$NH$_3$ in the southern hemisphere), although this may simply be related to degeneracies with the retrieval of atmospheric temperature.  On Saturn, the lower spatial resolution and signal-to-noise ratio made it harder to extract spatially variable parameters, but zonal-mean retrievals of temperatures revealed the band of emission at 40$^\circ$N remaining from Saturn's 2010-2011 springtime storm, some 3-4 K warmer than the background atmosphere \citep[consistent with ][]{14achterberg}.  Furthermore, Saturn's zonal PH$_3$ distribution showed the same equatorial enhancement as previous Cassini studies.  Zonally resolved spectra did not have sufficient signal to constrain latitudinal variations of Saturn's $^{15}$N/$^{14}$N ratio. 

The primordial $^{15}$N/$^{14}$N ratios on both Jupiter and Saturn imply that cosmochemical reservoirs of $^{15}$N-enriched ammonia could not have contributed significantly to their accretion.  The results favour primordial N$_2$ as the dominant contributor to the giant planet nitrogen inventory, accreted either in the gas phase or as ices formed in cold environments at great heliocentric distances.  In the former case, although N$_2$ is expected to be the dominant form of nitrogen in the solar nebula, some mechanism must be found to explain the supersolar N/H ratios on both planets, while keeping C/N solar on Jupiter and supersolar on Saturn.  In the latter case, we must explain how these ices of a distant origin could have survived transport to the forming protoplanets without significant devolatilisation.   Saturn's $^{15}$N/$^{14}$N ratio and C/N ratio must be confirmed via \textit{in situ} sampling \citep{14mousis,09marty}, but we hope that this upper limit will prove to be a useful constraint for modelling the chemical inventory and evolution of our giant planets.


\section*{Acknowledgments}
Fletcher was supported by a Royal Society Research Fellowship at the University of Oxford.  Fletcher, Greathouse and Orton were visiting astronomers at the Infrared Telescope Facility, which is operated by the University of Hawaii under Cooperative Agreement no. NNX-08AE38A with the National Aeronautics and Space Administration, Science Mission Directorate, Planetary Astronomy Program.  The UK authors acknowledge the support of the Science and Technology Facilities Council (STFC).  Mousis acknowledges support from CNES.  A portion of this work was performed by Orton at the Jet Propulsion Laboratory, California Institute of Technology, under a contract with NASA.  We are extremely grateful to John Lacy and Matt Richter for their assistance in understanding the performance of the TEXES instrument and the uncertainties related to calibration, and to Tobias Owen for helpful discussions.  We thank Thierry Fouchet and one anonymous reviewer for their helpful critiques of the manuscript.

\bibliographystyle{elsarticle-harv}
\bibliography{references_master}







\end{document}